\documentclass[showpacs,aps,reprint,superscriptaddress,showkeys,floatfix,citeautoscript]{revtex4-1}

\usepackage{bm}
\usepackage[pdftex]{graphicx}
\usepackage{amsmath}
\usepackage{amsfonts}
\usepackage{amssymb}
\usepackage{epstopdf}
\usepackage{color}

\usepackage{dsfont}
\usepackage{bm}
\usepackage[normalem]{ulem}
\usepackage{hyperref}
\hypersetup{
    colorlinks,%
    citecolor=blue,%
    linkcolor=blue,%
    urlcolor=blue
}

\usepackage[dvipsnames]{xcolor}

\newcommand{\mueV}{~\ensuremath{\mu}\text{eV}}
\newcommand{\mum}{~\ensuremath{\mu}\text{m}}

\begin{document}

\newcommand{\be}   {\begin{equation}}
\newcommand{\ee}   {\end{equation}}
\newcommand{\ba}   {\begin{eqnarray}}
\newcommand{\ea}   {\end{eqnarray}}
\newcommand{\ve}  {\varepsilon}

\newcommand{\state} {\mbox{\scriptsize state}}
\newcommand{\band} {\mbox{\scriptsize band}}
\newcommand{\Dis} {\mbox{\scriptsize dis}}
\newcommand{\vareps}{\varepsilon}
\newcommand{\tGamma}{\tilde{\Gamma}}


\title{ Conductance and Kondo Interference beyond Proportional Coupling }
\author{Luis G.~G.~V. Dias da Silva}
\affiliation{Instituto de F\'{\i}sica, Universidade de S\~{a}o Paulo,
C.P.\ 66318, 05315--970 S\~{a}o Paulo, SP, Brazil}
\author{Caio H. Lewenkopf}
\affiliation{Instituto de F\'{\i}sica, Universidade Federal Fluminense, 
24210-346 Niter\'oi, Brazil}
\author{Edson Vernek}
\affiliation{Instituto de F\'{\i}sica, Universidade Federal de Uberl\^andia, 
Uberl\^andia, Minas Gerais 38400-902, Brazil.}
\author{Gerson J. Ferreira }
\affiliation{Instituto de F\'{\i}sica, Universidade Federal de Uberl\^andia, 
Uberl\^andia, Minas Gerais 38400-902, Brazil.}
\author{Sergio E. Ulloa}
\affiliation{Department of Physics and Astronomy, and Nanoscale
and Quantum Phenomena Institute, Ohio University, Athens, Ohio
45701-2979, USA}

\date{\today}

\begin{abstract}

The transport properties of nanostructured systems are deeply affected by the geometry of the effective connections to metallic leads. In this work we derive a conductance expression for a class of interacting systems whose connectivity geometries do not meet the Meir-Wingreen proportional coupling condition.  
As an interesting application, we consider a quantum dot connected coherently to tunable electronic cavity modes.
The structure is shown to exhibit a well-defined Kondo effect over a wide range of coupling strengths between the two subsystems.  In agreement with recent experimental results, the calculated conductance curves exhibit strong modulations and asymmetric behavior as different cavity modes are swept through the Fermi level. These conductance modulations occur, 
however, while maintaining robust Kondo singlet correlations of the dot with the electronic reservoir, a direct consequence of the lopsided nature of the device.
\end{abstract} 

\maketitle

The quantum coupling of spatially localized discrete levels to cavity modes has emerged as a key tool for 
quantum information processing in different contexts, from cavity systems in atoms \cite{Raimond:Rev.Mod.Phys.:565--582:2001} 
and semiconductor quantum dots \cite{Hennessy:Nature:896--899:2007}  to  exciton-polariton condensates in optical 
systems  \cite{Kasprzak:Nature:409--414:2006}. 
Similarly, coherent coupling of electronic modes to discrete quantum systems has been explored in quantum 
corrals created on metallic surfaces \cite{Crommie}, allowing the manipulation and control of quantum information 
over regions a few nanometers across \cite{Manoharan}.
Recent experiments have extended this fascinating line of inquiry to systems implemented  on two-dimensional 
electronic structures in semiconductors \cite{Klaus2015,Sellier2016}. These new systems have paved the way 
for quantum engineering in integrated, scalable nanoscale systems  with great flexibility on geometries 
and interesting physical behavior.

The control of quantum dot (QD) characteristics in these systems, such as the tunnel coupling  to external current leads, have 
also allowed the experimental study of the Kondo regime, an emblematic many-body effect \cite{Goldhaber-Gordon:156:1998,Cronenwett:540:1998}. 
In this regime, 
the net magnetic moment of an unpaired spin in  the QD becomes effectively screened by the conduction 
electrons in the leads, forming a delocalized quantum singlet that involves correlations with the electronic spins
in the lead reservoirs \cite{HewsonBook}. 
Moreover, the coupling of a QD to reservoirs 
with non-trivial energy dependence gives rise to a variety of interesting effects on the ensuing Kondo state, 
including the appearance of zero-field splittings of the Kondo resonance \cite{KondoBox,Silva2006,Silva2008}.
As QD systems are designed to interact with increasingly complex structures, one is led to ask how such many-body 
correlations would evolve.

The standard theoretical tool for the description of the two-terminal conductance through interacting regions is the Meir-Wingreen (MW) generalization of the Landauer formula for correlated  systems \cite{MeirWingreen92}.  The MW expression is particularly useful in cases where the coupling matrix elements between the leads and the system 
are related 
to each other by a multiplicative factor. 
This condition was later dubbed ``proportional coupling" (PC) \cite{Meir1993} and it is essential 
in writing the conductance in terms of the system's retarded Green's function. 
In many cases, however, the PC description is inadequate \cite{Affleck2013} and the evaluation 
of the conductance requires an alternative treatment. 

A remarkable example of a nanoscale device with non-PC geometry was recently investigated 
 in Ref. \cite{Klaus2015}. 
They demonstrated coherent coupling between a QD in the Coulomb blockade  regime and a larger,  cavity-like region inscribed electrostatically onto the same two-dimensional  electron 
gas (2DEG). The QD is coupled to two metallic leads while the cavity itself is coupled 
to only  one of them, clearly breaking the PC condition.  
The size of the cavity and its coupling to the QD can be controlled by gate 
voltages on the device, allowing for fine control over the 
spacing between cavity resonances, the tunnel rate of electrons between cavity and QD,
and the dot-cavity coupling over a wide range, while studying the conductance of the 
entire structure.

In this paper we 
extend the applicability of the MW expression 
to a large class of non-PC cases, providing theoretical tools to analyze the transport properties and temperature dependence of
systems with a single interacting level (such as a QD) embedded in complex structures, as some studied recently
\cite{Klaus2015,Sellier2016}.
We find it is possible to write the linear
conductance
of such systems
as
\begin{equation}
\label{eq:MainConductance}
G = \frac{2 e^2}{\hbar} 
\frac{ \tGamma_L(\ve_F)  \tGamma_R(\ve_F)}
{\tGamma_L(\ve_F) + \tGamma_R(\ve_F)}
 \int\! d\omega \,  
 \left(- \frac{\partial f_0}{\partial \omega}\right)  
A_{d} (\omega),
\end{equation}  
where $f_0$ is the equilibrium Fermi function, the couplings ${\tGamma}_{L,R}(\ve_F)$ 
are effective hybridization functions to left ($L$) and right ($R$) leads, 
$A_{d} (\omega)=(-1/\pi)  \mbox{Im} \, G_{d}^r (\omega)$ 
the spectral function, and $G^r_{d}$ is the retarded Green's function at the QD. \@
The latter two functions can be accurately calculated through a variety of 
techniques, such as Wilson's numerical renormalization group (NRG) \cite{NRGRMP2008}.

Although deceptively similar to the MW conductance formula for a single-level QD \cite{Meir1993}, this expression incorporates the connection of the entire complex system to each lead through the effective hybridization functions  ${\tGamma}_{L,R}(\ve_F)$.
A crucial difference is that, in the original formula \cite{MeirWingreen92}, the hybridization is represented by 
matrices of functions
$\mathbf{\Gamma}^{L,R}$ 
involving the couplings and the density of states in the leads. Here, such complexities are encoded 
in the intricate energy structure of ${\tGamma}_{L,R}(\omega)$. 
As we will see below, these functions can be obtained after careful consideration of the effective connectivity of the system.

Next, we use this approach to successfully describe and provide further insight on conductance 
measurements of a QD coupled to a cavity \cite{Klaus2015}.  
We implement a realistic model of the curved electrostatic reflector used to 
define the cavity in experiments, utilizing both analytical
and numerical approaches.  
We further calculate the QD spectral density required by Eq.\ \eqref{eq:MainConductance} by 
applying NRG to an effective Anderson model that incorporates the cavity.
Our results show contrasting transport properties in the weak- 
and strong-coupling regimes, in excellent agreement with experiments.  
As the coupling to the cavity sets in, the conductance is strongly modulated, especially as 
different cavity resonances are swept 
through the Fermi level in the leads by applied gates \cite{Klaus2015}.  
Moreover, the NRG calculations  allow us to relate the 
conductance behavior to other intrinsic characteristics, such as
the Kondo temperature $T_K$.  We find that even as the 
conductance peaks are strongly distorted due to the interaction with the cavity modes, the Kondo 
screening remains robust, with larger $T_K$ values for stronger cavity coupling. 

\paragraph*{MW formula beyond proportional coupling.}
Proportional coupled systems are those in which the 
coupling matrices of the interacting system to $L$ and $R$ 
 leads are proportional to each other, namely, 
$\mathbf{\Gamma}^{R}(\omega) = \lambda\mathbf{\Gamma}^{L}(\omega)$
where $\lambda$ is a  constant factor \cite{MeirWingreen92}. 
This condition is clearly violated in the case of a QD connected to a cavity on only one lead, 
such as in Fig.\ 
\ref{fig:Cavity}. 
An electron in the dot is transmitted from $L$ 
by a direct tunneling process regulated by 
the coupling matrix element $V_{dL}$ and the density of states in that lead. 
In contrast, the transmission to the right 
involves the coherent interference between multiple paths 
that include the cavity resonances and states in 
$R$.
Figure 1(b) indicates the different 
dot-lead ($V_{dR}$), and cavity-lead ($V_{cR}$) couplings
that 
enter as non-zero elements in $\mathbf{\Gamma}^R$, while the cavity-lead 
couplings are zero in $\mathbf{\Gamma}^L$, thereby
making 
the system evidently non-proportional \cite{Supp}.

\begin{figure}[ht!]
  \centering
\includegraphics[width=\columnwidth,keepaspectratio=true]{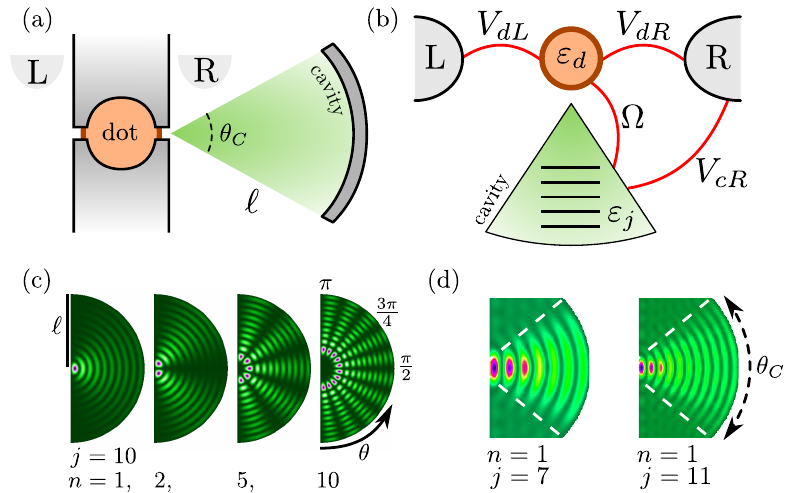}
  \caption{
 (a) Experimental dot+cavity system; the cavity has radius $\ell$ and 
 aperture $\theta_C$.
 (b) Schematic of single-level dot ($\varepsilon_d$) coupled to multi-mode cavity ($\varepsilon_j$). 
 The dot is connected to leads ($L$ and $R$), while cavity is only coupled to the $R$-lead; 
 coupling matrix elements are indicated. 
 (c) Cavity modes for $\theta_C = \pi$ are described by Bessel modes $\psi_{n,j}(r,\theta)$. The $n=1$ 
 modes dominate the LDOS at $r\approx 0$.
 (d) Kwant mode simulation for finite aperture cavity ($\theta_C = \pi/2$) coupled to wide leads shows good 
 agreement with Bessel modes.
 }
 \label{fig:Cavity}
\end{figure}

The main technical difficulty in obtaining a transport formula is the calculation of the lesser Green's functions matrix
$\mathbf{G}^{<}$ for the interacting region, which appears in the general expression for the current  \cite{MeirWingreen92}.
The latter gives the current through the $L$ ($R$) lead as
\begin{align}
\label{eq:MeirWingreen}
J_{L(R)} =  \frac{ie}{h} \int d\omega \, & \mbox{tr} \left( \mathbf{\Gamma}^{L(R)} (\omega)\Big\{ \mathbf{G}^<(\omega)  \right. \nonumber \\ 
& \left. + f_{L(R)}(\omega) \left[ \mathbf{G}^r(\omega) - \mathbf{G}^a(\omega) \right]\Big\} \right) \, ,
\end{align}
where $\mathbf{G}^{r(a)}$ is the retarded (advanced) Green's function matrix \cite{Supp} 
and $f_{L(R)}$ is the Fermi distribution at the $L (R)$ lead with chemical potential $\mu_{L(R)}$.  
Proportional coupling and current conservation make possible to simplify the calculation by ingeniously 
writing $J_{L(R)}$ in terms of $\mathbf{G}^{r(a)}(\omega)$. In contrast, for interacting non-PC systems 
away from equilibrium, the elimination of $\mathbf{G}^<$ is in general not possible. 
However, in the linear response regime it can be achieved by recalling that \cite{Affleck2013}
\begin{align}
\label{eq:linear_response}
\mathbf{G}^<(\omega) \approx \mathbf{G}^<_{\rm eq}(\omega) - \frac{\partial f_0}{\partial \omega} \mathbf\Pi(\omega) \Delta \mu 
+ O(\Delta \mu^2)  \, ,
\end{align} 
where $\Delta \mu = \mu_L - \mu_R$ and $\mathbf\Pi(\omega)$ has a slow $\omega$ dependence within energy 
windows of $k_BT$ corresponding to the experiments of interest.
These conditions eventually lead to
Eq.\ \eqref{eq:MainConductance}; the detailed derivation is provided in the supplement \cite{Supp}.
Notice that the 
structure of the system may result in a cumbersome derivation
of the 
${\tGamma}_{L,R}(\omega)$ entering 
Eq.\ \eqref{eq:MainConductance}.  We now specify the QD-cavity model that exemplifies this treatment. 

\paragraph*{Resonant cavity modes.}
The key experimental element is a ``mirror" that
focuses resonant modes onto the QD, both elements electrostatically defined on a 2DEG. The cavity has a length $\ell \sim 2$$\mu$m 
and angular aperture $\theta_C \sim 45^\circ$, as indicated in 
Fig.~\ref{fig:Cavity}(a).  Assuming circular symmetry, 
the normal modes are given by Bessel functions, $\psi_{n,j}(r,\theta) \! \simeq \! J_n(k_{n,j}r) \sin (n\theta)$. 
The dot-cavity coupling is maximal for modes with largest 
amplitude in the vicinity of $r\approx 0$, and dominated by resonances with $n=1$, 
given that $J_n(k r) \! \propto \! (kr)^n$ for $kr \ll 1$.
These modes have a characteristic energy spacing $\delta_{cav} \approx 200~\mu$eV for a cavity
with these dimensions, in agreement 
with the resonance separations in the experiment \cite{Klaus2015}  and confirmed by Kwant calculations \cite{Supp,kwant}.

It is remarkable that although the cavity is immersed in the $R$-lead, it can be
tuned to produce sharply peaked resonances that 
strongly modify $\tGamma_R(\omega)$, providing different electronic paths for
the current. 
In the experiment, a gate voltage shifts the cavity resonance levels and the coupling to the QD.\@  
This tunability can be 
incorporated in the interacting QD  model as follows.  

\paragraph*{Interacting quantum impurity model.}
The Hamiltonian for this system can be written as 
$H=H_{\rm dot} + H_{\rm cavity}+ H_{\rm leads} + H_{\rm coupling}$,
where
\begin{eqnarray} \label{eq:Hamiltonian}
H_{\rm dot} &=& \sum_{\sigma} \ve_d^{} c^{\dagger}_{d\sigma} 
c^{\phantom{\dagger}}_{d\sigma} + U n_{d\uparrow} n_{d\downarrow},\\ 
 H_{\rm cavity} &=& \sum_{j,\sigma}\ve_j^{}  a^{\dagger}_{j\sigma} 
a^{\phantom{\dagger}}_{j\sigma}, \\
 H_{\rm leads} &=&   \sum\limits_{\alpha,\mathbf{k},\sigma} 
\ve_{\alpha\mathbf{k}} ^{}\, c^{\dagger}_{\alpha\mathbf{k}\sigma}
  c^{\phantom{\dagger}}_{\alpha\mathbf{k}\sigma} \, .
\end{eqnarray}
Here $c^{\dagger}_{d\sigma}$, $a^{\dagger}_{j\sigma}$, and  $c^{\dagger}_{\alpha\mathbf{k}\sigma}$
create a spin-$\sigma$ electron in the dot, the $j$th mode of the cavity, and each of the leads $\alpha=L,R$.
The resonances are assumed equally spaced, $\ve_j=\epsilon_c+(j-1)\delta_{cav}$, where 
$\epsilon_c$ is shifted by a gate voltage; leads
have a flat density of states
$\rho(\omega)=\rho_0 \Theta(D-|\omega|)$, symmetric about the Fermi energy
($\omega=0$). For simplicity all
couplings are assumed local, real and independent of either momentum in the leads 
or cavity-mode index $j$. The coupling Hamiltonian is then,
see Fig.\  \ref{fig:Cavity}(b),
\begin{align}
  H_{\rm coupling} = &
  \sum\limits_{\alpha,\mathbf{k},\sigma} V_{d\alpha}^{}  \, 
c^{\dagger}_{d\sigma} c^{\phantom{\dagger}}_{\alpha\mathbf{k}\sigma} + 
  V_{cR}^{} \sum\limits_{j,\mathbf{k},\sigma} a^{\dagger}_{j\sigma} 
c^{\phantom{\dagger}}_{R\mathbf{k}\sigma} \nonumber \\
& +\Omega\sum_{j,\sigma} c^{\dagger}_{d\sigma}  
a^{\phantom{\dagger}}_{j\sigma} + \mathrm{H.c.} \, . 
\end{align}

\paragraph*{QD effective decay widths.}
As the Coulomb interactions are localized in the QD, one can find 
its effective couplings to $L$ and $R$ leads and the
cavity, by calculating the dot retarded Green's function for the system with $U\!=\!0$, 
$G^{(0),r}_d (\omega^+)\! \equiv \!  \langle\!\langle c^{\phantom{\dagger}}_{d\sigma} 
;c^{\dagger}_{d\sigma} \rangle\!\rangle_{\omega}$.
In the wide-band limit for the leads, 
$\sum_{\mathbf{k}} (\omega^+ -\ve_\mathbf{k})^{-1} \rightarrow -i\pi \rho_0$, 
we obtain 
$G^{(0),r}_d(\omega) = 
({\omega - \ve_d - \Sigma^{(0)}_d(\omega)})^{-1} $, 
where 
\begin{equation}\label{eq:SelfEnergy0dot}
 \Sigma^{(0)}_d(\omega) = - \frac{i}{2}\left( \Gamma_{dL} +
\Gamma_{dR} \right) + \left( \Omega - \frac{i}{2} \sqrt{\Gamma_{dR}\Gamma_{cR}} 
\right)^2\tilde S(\omega) \, ,
\end{equation}
is the non-interacting self-energy.  Here, $\Gamma_{(c,d)\alpha}\equiv 2\pi \rho_0 
|V_{(c,d)\alpha}|^2$, for $ \alpha =L,R$, with the cavity structure contained in $S(\omega) \equiv 
\sum_{j} ({\omega - \ve_j})^{-1}$ and $\tilde{S}(\omega)=S(\omega)\left( 1 + i 
S(\omega) \Gamma_{cR}/2 \right)^{-1}$. 
The hybridization function of the (non-interacting) dot with the 
effective fermionic system is given by $\Delta(\omega)=-\mbox{Im} \Sigma^{(0)}_d(\omega)$. 
This approach can be extended to the 
interacting Green's function \cite{Silva2006,Silva2008}, as long as the 
interactions are restricted to the QD.

The interference of cavity modes and states in the leads is contained in the structure
of $\Delta(\omega)$, which yields a highly structured density of states of the ``effective"
Fermi reservoir in which the QD is embedded \cite{Supp}. 
Most importantly, the structure in $\Delta(\omega)$ affects strongly the Kondo state in the 
system once interactions set in. 
$\Delta(\omega)$ reliably
describes the experimental system once cavity parameters are extracted either from
a microscopic model, and/or determined from experiments 
\footnote{ 
The structure studied in Ref. \cite{Klaus2015} has charging 
energy $U\!=\!700 \,\mu$eV; dot-source (left)  and 
dot-drain (right)  couplings $\Gamma_{dL}\!\approx\!\Gamma_{dR} \!\approx\! 87 
\,\mu$eV; cavity broadening 
$\Gamma_{cR}\!\approx\! 40 \,\mu$eV; and cavity mode spacing $\delta_{cav}\!\approx\! \,220 
\mu$eV.\@  
Using $U\!=\!0.5D\approx\! 700\, \mu$eV, the NRG scales are then set as 
$\Gamma_{dL}\!=\!\Gamma_{dR}=0.125 U$, 
$\Gamma_{cR}\!=\! 0.06 U$ and $\delta_{cav}\! =\! 0.32 U$.}.

\paragraph*{Conductance for the interacting system.}
Eq.\ \eqref{eq:MainConductance} determines the conductance
through the system under different cavity+QD coupling regimes. 
The QD coupling to the left (source) reservoir is simply
$\tGamma_L = \Gamma_{dL}$. In contrast, the coupling to the right (drain) reservoir 
requires the full Green's function and results in \cite{Supp} 
\begin{multline}
\label{eq:DeltaR}
\tGamma_R(\omega) = \Gamma_{dR} + \Gamma_{cR}  \left|\tilde 
S(\omega)\right|^2 \left( \Omega^2 + \frac{\Gamma_{cR} \Gamma_{dR}}{4}\right)
+\\
\sqrt{\Gamma_{dR}\Gamma_{cR}} \,\, \tilde S(\omega)  \left( \Omega  - \frac{i}{2} 
\sqrt{\Gamma_{cR} \Gamma_{dR}} \right) + \text{H.c.} \, .
\end{multline}
This expression encodes information about all non-trivial interference processes taking place during
transport. The energy dependence of $\tGamma_R(\omega)$ prevents the use of the PC simplification, demanding the more general approach we put forward here. 
The spectral function needed in 
Eq.\ \eqref{eq:MainConductance} is obtained by an NRG approach that uses the full intricate structure
of the effective hybridization function $\Delta (\omega)$ coupling the interacting QD to the environment.

\begin{figure}[t]
\begin{center}
\includegraphics[width=1.0\columnwidth]{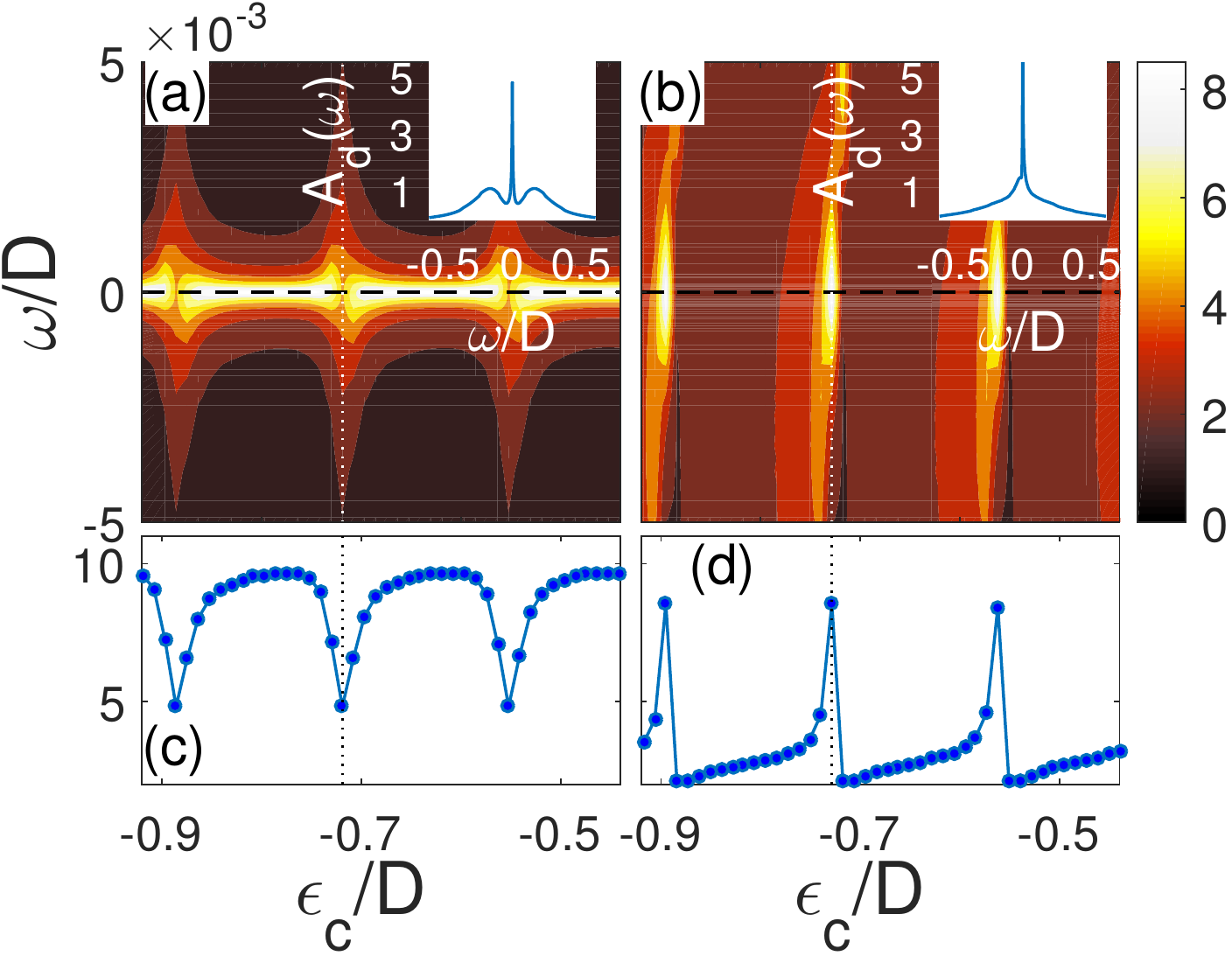}
\caption{NRG-calculated dot spectral density $A_d(\omega)$  for cavity gate 
voltages $\epsilon_c$ in the weak  [$\Omega=0.01D$, (a), (c)] and strong 
coupling [$\Omega=0.15D$, (b),(d)] regimes. Panels (c) and (d) show $A_d(0)$ vs $\epsilon_c$ (cuts through the horizontal dashed lines) 
Peaks in $A_d(0)$ correspond to dips in $\Delta(0)$ and vice versa \cite{Supp}. Insets show typical Kondo peaks in $A_d (\omega)$, present even when cavity modes dominate $\Delta(0)$ (vertical dotted lines).  
} 
\label{fig:SpecDens_Contour}
\end{center}
\end{figure}

Before discussing the conductance, we analyze the QD spectral function.
In general, $A_d(\omega)$ shows a sequence of asymmetric features whenever $\epsilon_c$ shifts cavity modes
near the Fermi level ($\omega\!=\!0$), with characteristic shape and width that changes strongly
with coupling $\Omega$.
Figure \ref{fig:SpecDens_Contour} illustrates this behavior for weak ($\Omega<\Gamma_{cR}/2$) and 
strong ($\Omega>\Gamma_{cR}/2$) dot-cavity coupling regimes.
For weak coupling [Fig.~\ref{fig:SpecDens_Contour}(a)\&(c)], 
the modulation is marked by diagonal ``valleys" whenever a cavity mode contributes to $\Delta(0)$, 
separated by bright peaks in $A_d$.
The large $\Omega$ regime [Fig.~\ref{fig:SpecDens_Contour}(b)\&(d)] is drastically different: $\Delta(0)$ exhibits Fano asymmetric lineshapes as a function of $\epsilon_c$, leading to  sharp asymmetric peaks in $A_d(\omega\!<\!T_K)$ \cite{Supp}.

This behavior can be qualitatively understood in terms of the Friedel sum rule (FSR)
\cite{Silva2006,Vaugier:Phys.Rev.B:165112:2007,Silva2013}, as 
$A_d(0)$ is inversely  proportional to 
$\Delta(0)$.  
Accordingly, when a resonant peak of $\Delta(\omega)$ lies close to the Fermi energy, it causes 
a downturn in the spectral function, and a consequent {\em splitting} of the Kondo peak may appear in $A_d$ in the $\omega < T_K$ range \cite{Silva2006}. 
Such splittings do appear for some $\epsilon_c$ values, where $A_d$ shows two local maxima away from the $\omega\!=\!0$ mark in Fig.~\ref{fig:SpecDens_Contour}(b) (see details in  \cite{Supp}).
Nonetheless, even at these points, $A_d(\omega)$ shows fully-developed Kondo resonances of width $\sim T_K$ in between Hubbard peaks (insets in Figs.~\ref{fig:SpecDens_Contour}(a) and (b)).

\begin{figure}[t!]
\begin{center}
\includegraphics[width=1.0\columnwidth]{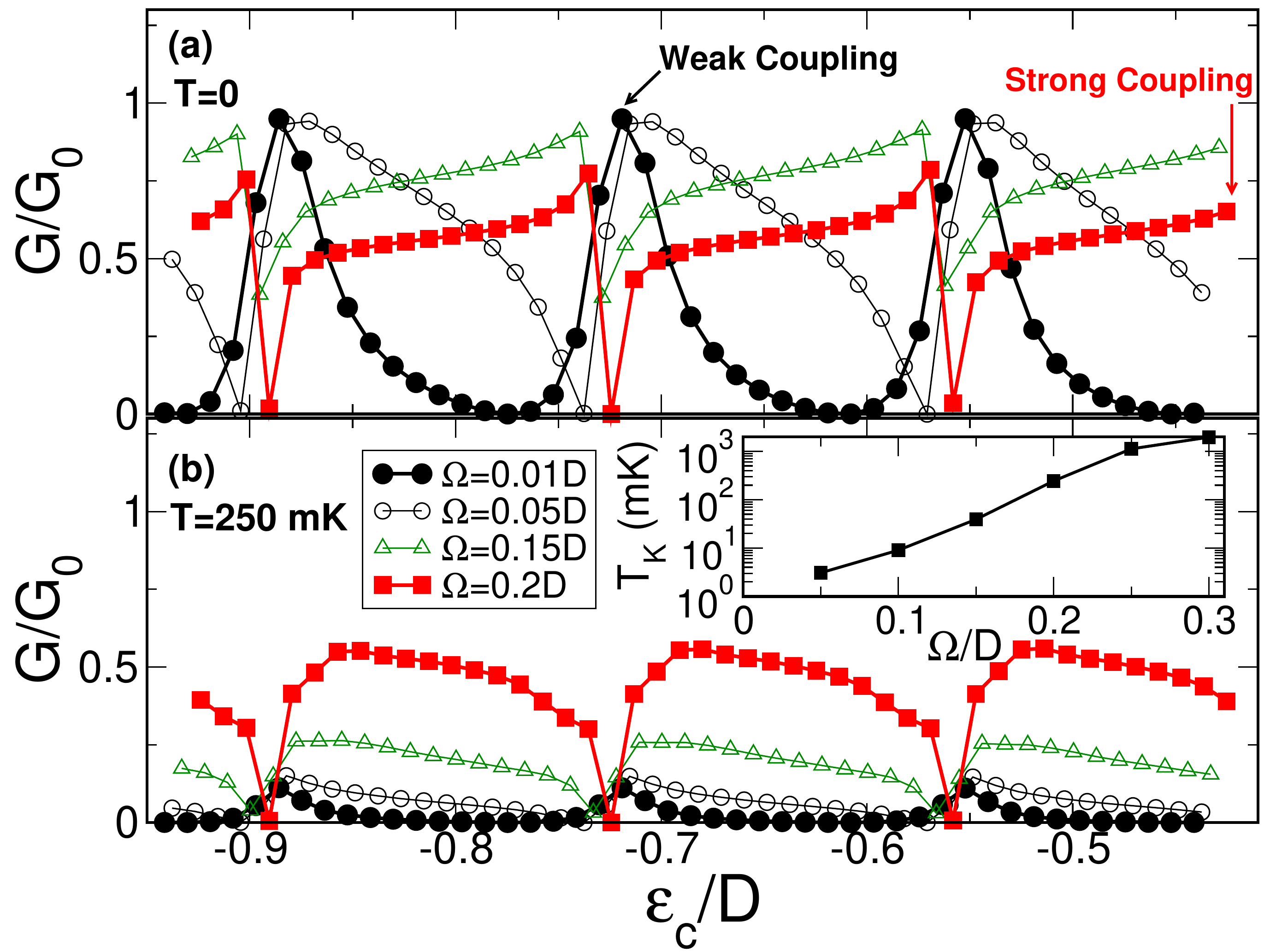}
\caption{Conductance  $G/G_0$ versus cavity gate voltages $\epsilon_c$ with cavity-dot couplings ranging from
the weak ($\Omega\!=\!0.01D$) to the strong coupling regime ($\Omega\!=\!0.2D$) for (a) $T\!=\!0$ and (b) $T\!=\!0.031U$ (or $T\!=\!250$mK for $U\!=\!0.7$ meV). 
Inset: Kondo temperature as a function of cavity-dot coupling $\Omega$ 
for $\epsilon_c=-0.9D$.}
\label{fig:Conductance}
\end{center}
\end{figure}

The resulting conductance $G$ (in units of $G_0=2e^2/h$) 
is shown in Fig.\ \ref{fig:Conductance} vs cavity voltage 
$\varepsilon_c$, for $\Omega$ values from $0.01D$ (weak) 
to  $0.2D$ (strong coupling) and for $T=0$ \& 250mK.\@ 
At low temperatures and small $\Omega$, the conductance exhibits 
a quantized peak whenever a cavity resonance is near the Fermi level, in agreement with the experimental result \cite{Klaus2015}.  The conductance drops away 
with $\epsilon_c$ as destructive interference sets in and results in a
non-zero scattering shift associated with the strongly asymmetric $A_d(\omega)$, as expected from 
the FSR.\@ 
Conversely, when a cavity resonance is aligned with the Fermi level in the strong coupling regime, a Fano-like {\em dip}
appears in the conductance, 
with a width much smaller than the cavity level spacing. This feature is also consistent with the experimental data of Ref.\ \cite{Klaus2015}.
Finite temperatures do not result in qualitative changes of this picture, but suppress the magnitude of $G$, as one would expect, with a larger effect for $T_K$ values below
the temperature of the reservoir (here 250mK).

Notice that the spinful QD remains in the Kondo regime over this range of coupling to the cavity.
In fact, the Kondo screening is stronger for larger $\Omega$, as monitored by the value of $T_K$.
To quantify this, we
calculate $T_K$ from the magnetic susceptibility curves obtained from NRG, a procedure that
focuses on how the Kondo fixed point is reached at lower energies, and does not rely on the behavior
of the spectral density \cite{NRGRMP2008}.
The inset in Fig.\ \ref{fig:Conductance} shows $T_K$ increasing rapidly with larger  QD-cavity coupling $\Omega$. 
For $\Omega=0.15D-0.2D$, we obtain  $T_K \sim 0.0048U-0.03U$; with the experimental $U=0.7$meV, this translates into $T_K \sim 40-240$mK, which is consistent with the observed  value of $\sim100$mK, obtained from the conductance peak width (see supplement in \cite{Klaus2015}). 
Our calculations also show $T_K$ to depend weakly on $\epsilon_c$. 
This might appear counterintuitive, as $\Delta(0)$ is strongly modulated by changes 
in $\epsilon_c$, but the explanation is simple: 
The effective coupling defining the Kondo temperature (e.g., $\Gamma$ in Haldane's expression 
\cite{Haldane1978}) is given not by $\Delta(0)$, 
but rather by an integral over the full bandwidth, $\Gamma \!\propto\!{\int 
\Delta(\omega) d(\omega/D)}$ \cite{Gonzalez-Buxton98}.  
This ``$\Gamma$" depends strongly on the dot-cavity coupling $\Omega$ (thereby giving the strong 
variation of $T_K$ with $\Omega$) while only weakly with $\epsilon_c$, whose main effect is to shift the peaks in $\Delta(\omega)$.

The increasing $T_K$ indicates that the screening of the QD spin by the composite cavity-lead environment 
is in fact more robust for larger $\Omega$, which is confirmed by an NRG analysis of the thermal properties of the QD.
This is remarkable behavior, as the strong variation in $A_d(\omega)$ and 
resulting conductance are drastically different from the simply-connected QD in the Kondo regime. 

\paragraph*{Discussion.}
We have presented an approach that allows one to calculate the   
linear conductance through interacting systems
beyond the proportional coupling approximation. 
This opens the possibility of studying interesting systems with complex geometries where quantum interference introduces 
non-trivial energy dependence on the effective decay widths ${\tGamma}_\alpha$. We have illustrated the power of the method by analyzing a recent experiment with very interesting geometry \cite{Klaus2015}. 
Despite the observed splitting and strong modulation of conductance
peaks for growing cavity coupling, we find that the Kondo screening is in fact strengthened, as 
characterized by a larger $T_K$.   
This interpretation is supported by calculations of the conductance in excellent agreement with experiment.  
It would be interesting to be able to measure the expected phase shifts introduced by the interaction with the cavity
to provide further insights into the coherent interference that these many-body coupled systems experience.

We acknowledge useful discussions with C. R\"ossler, T. Ihn, K. Ensslin, and N. Sandler.  
LDS\ acknowledges support from CNPq grants 307107/2013-2 and 449148/2014-9, PRP-USP 
NAP-QNano and FAPESP grant 2016/18495-4. 
SEU received support from NSF grant DMR 1508325, and the Aspen Center for Physics, NSF grant PHY-1066293. 
CHL is supported by CNPq grant 308801/2015-6 and FAPERJ grant E-26/202.917/2015. GJF and EV acknowledge financial support from CNPq and FAPEMIG.

%

\widetext
\clearpage
\setcounter{equation}{0}
\setcounter{figure}{0}
\setcounter{table}{0}
\setcounter{page}{1}
\makeatletter
\renewcommand{\theequation}{S\arabic{equation}}
\renewcommand{\thefigure}{S\arabic{figure}}
\renewcommand{\bibnumfmt}[1]{[S#1]}
\renewcommand{\citenumfont}[1]{S#1}


\begin{center}
\textbf{\large Supplemental Material for \\ \textit{ Conductance and Kondo Interference Beyond Proportional Coupling}}
\end{center}


\section{System geometry and model parameters}

The model parameters we use in this paper have been inferred from the experimental 
data from Ref.~\onlinecite{Supp::Klaus2015} combined with analytical estimates and numerical calculations. Here, 
we provide more details on the numerical simulations. 

The experimental setup of Ref.~\onlinecite{Supp::Klaus2015} consists of a cavity 
focusing resonant modes into a quantum dot, both set on a GaAs two-dimensional electron gas (2DEG). The cavity 
has a radius $\ell \sim 2\mum$ and an angular aperture $\theta_C \sim 45^\circ$,
as indicated in Fig.~\ref{fig:Cavity}(a).
To obtain a simple, yet accurate description of the non-interacting modes of the cavity, we 
consider its eigenstates to be approximately given by Bessel functions.
The Bessel approximation becomes exact for a large aperture $\theta_C 
\rightarrow 180^\circ$, as  the cavity approaches a semi-circle shape 
[Fig.~\ref{fig:Cavity}(b)].
In the following we show that this approximation leads to a level spacing that 
agrees remarkably  well with the experimental \cite{Supp::Klaus2015} peak energy splitting 
$\delta_{cav} \approx 220\mueV$. 

\begin{figure}[ht!]
  \centering
  \includegraphics[width=0.6\columnwidth,keepaspectratio=true]{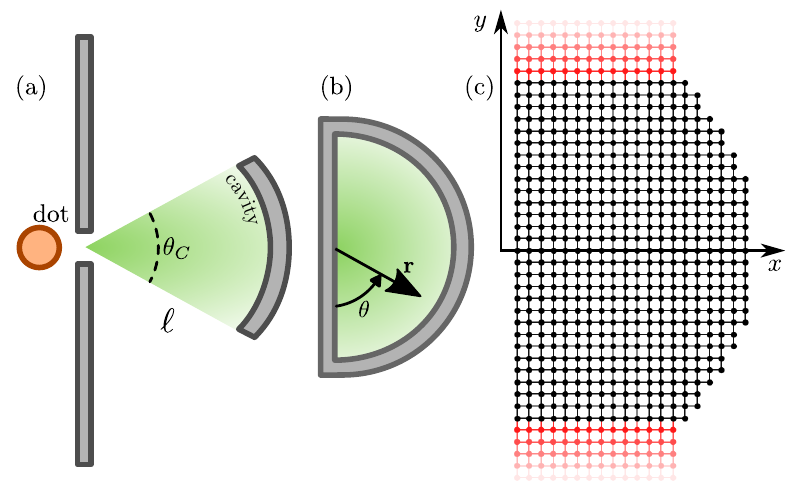}
  \caption{(a) Illustration of the dot-cavity coupled system indicating the 
radius $\ell = 2\mum$ and angular aperture $\theta_C$ of the resonant cavity. 
(b) Two-dimensional system of coordinates $\bm{r} = (r, \theta)$ for the 
semi-circle approximation for the cavity modes corresponding to the limit 
$\theta_C \rightarrow 180^\circ$. 
(c) Example of the finite differences lattice 
model with leads (in red) implemented in Kwant. In this illustration the grid 
step size is large ($\sim 100$~nm) for better visualization, while for the simulations the step size reduced ($\sim 2$~nm).
  }
  \label{fig:Cavity}
\end{figure}

 Assuming hard-wall boundary conditions, the solution 
for the Schr\"odinger equation in cylindrical coordinates results in eigenstates 
$\psi_{n,j}(r,\theta)$ given by Bessel functions $J_n(z)$, and eigenenergies 
$\varepsilon_{n,j}$ set by the $j^\text{th}$ zero $z_{n,j}$ of $J_n(z)$ at $r = 
\ell$, which reads
\begin{eqnarray}
 \psi_{n, j}(r, \theta) &=& C_{n,j} \sqrt{\frac{2}{\pi}}\sin(n\theta) J_n\left(k_{n,j} r\right),
 \label{eq:psibessel}\\
 \varepsilon_{n,j} &=& \frac{\hbar^2}{2m} \left(\frac{z_{n,j}}{\ell}\right)^2,
\end{eqnarray}
where $k_{n,j} = \sqrt{2m\varepsilon_{n,j}/\hbar^2}$, and $C_{n,j}$ is a normalization constant.
To satisfy the boundary condition at the linear wall of the semi-circle ($x=0$), the 
index $n = 1, 2, 3, 4, \dots$  must be a non-zero integer. 

Near the Fermi level $k_{n,j} \approx k_F = 2\pi/\lambda_F$, where $\lambda_F$ is the 
Fermi wavelength of the 2DEG under the resonant cavity. For $\ell \approx 2 \mum$
one gets $k_{n,j}\ell \approx 2\pi \ell/\lambda_F \gg 1$, which allow us to use 
the asymptotic limit of the Bessel functions\cite{abramowitz1964} to find 
analytical expression for the zeros $z_{n,j}$. Since $J_n(z) \approx \sqrt{2/\pi 
z}\cos(z - n \pi/2 - \pi/4)$ we find
\begin{equation}
 z_{n,j} = \frac{3\pi}{4} + \frac{\pi}{2}(n+2j) = \frac{3\pi}{4} + \frac{\pi}{2} l \equiv z_l.
\end{equation}
The $n$ and $j$ quantum labels become degenerate, and the Bessel zeros become 
simply $z_l$, with $l = (n+2j)$.  The integer $l$ is odd (even) whenever $n$ is 
odd (even). Consequently $\varepsilon_{n,j} \rightarrow \varepsilon_l$ near the 
Fermi level, 
\begin{equation}
 \varepsilon_l = \frac{\hbar^2}{2m \ell^2}\left(\frac{3\pi}{4} + \frac{\pi}{2}l\right)^2.
\end{equation}

\begin{figure}[ht!]
  \centering
  \includegraphics[width=0.6\columnwidth,keepaspectratio=true]{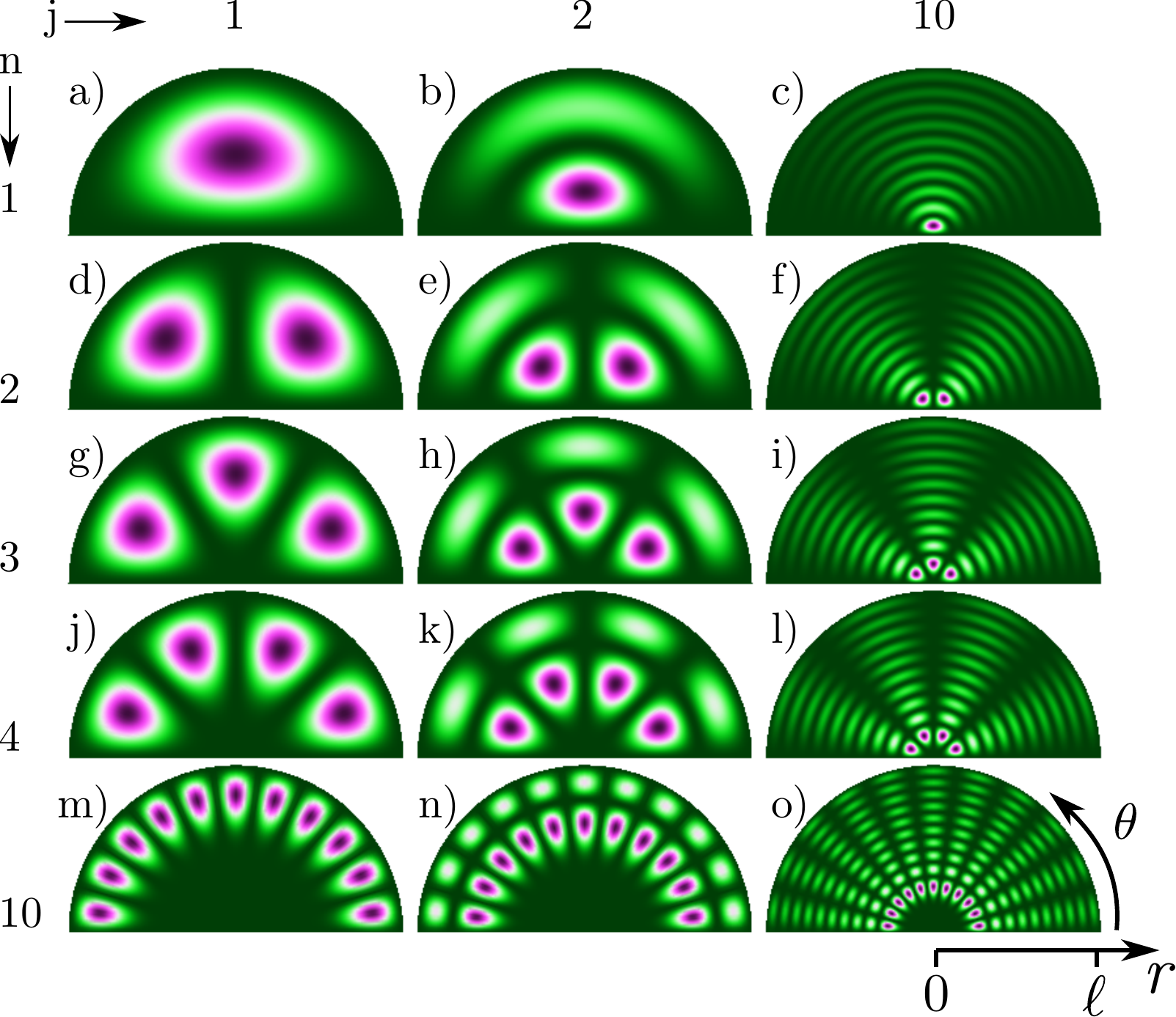}
  \caption{Bessel modes $|\psi_{n,j}(r,\theta)|^2 \propto \big[\sin(n\theta) J_n\left(k_{n,j} r\right)\big]^2$ 
  (see Eq.~\eqref{eq:psibessel}) representing the LDOS peaks 
  of the resonant  cavity of $\ell = 2\mum$. The panel lines correspond to $n = 1, 2, 
  3, 4$, and $10$, and the columns are for $j=1, 2$, and $10$, as indicated. Excluding the 
  boundaries, the number of nodes along $\theta$ is $n-1$, and along $r$ it is 
  $j-1$. The $n=1$ modes dominates the LDOS near $r = 0$, where the cavity effectively couples to the dot.
  }
  \label{fig:BesselDens}
\end{figure}

The coupling of the dot with the resonant modes of the cavity occurs via the 
split-gate set  by the linear electrodes in Fig.~\ref{fig:Cavity}(a). Therefore 
the relevant quantity is the LDOS $\propto |\psi_{n,j}(r,\theta)|^2$ of the 
cavity modes in the vicinity of this region, i.e. $\bm{r} \sim 0$. Figure 
\ref{fig:BesselDens} shows $|\psi_{n,j}(r,\theta)|^2$ for different $n$ and $j$. 
Since $J_n(kr) \propto (kr)^n$ for $kr \ll 1$, 
near $r = 0$ the dominant coupling must be given by $n=1$, 
yielding odd $l$. 

We conclude that the energy spacing between cavity resonant modes that 
are effectively coupled to the dot is
\begin{equation}
 \delta_{cav} = \varepsilon_{l+2} - \varepsilon_l = \frac{\hbar^2}{2m \ell^2} 
\frac{\pi^2}{2}(5+2l).
\end{equation}
Considering the experimental data of Ref.~\onlinecite{Supp::Klaus2015},  $\lambda_F = 
53$~nm and $\ell = 2\mum$, we obtain $\varepsilon_F \approx 8$~meV and $l \sim 150$ 
for $\varepsilon_l \sim \varepsilon_F$, corresponding to 75 even $l$ and 75 odd 
$l$ occupied resonant modes. From these we find $\delta_{cav} \approx 200\mueV$, 
which matches the experimental energy splitting between resonances reported in 
Ref.~\onlinecite{Supp::Klaus2015}.

\begin{figure}[ht!]
  \centering
  \includegraphics[width=0.6\columnwidth,keepaspectratio=true]{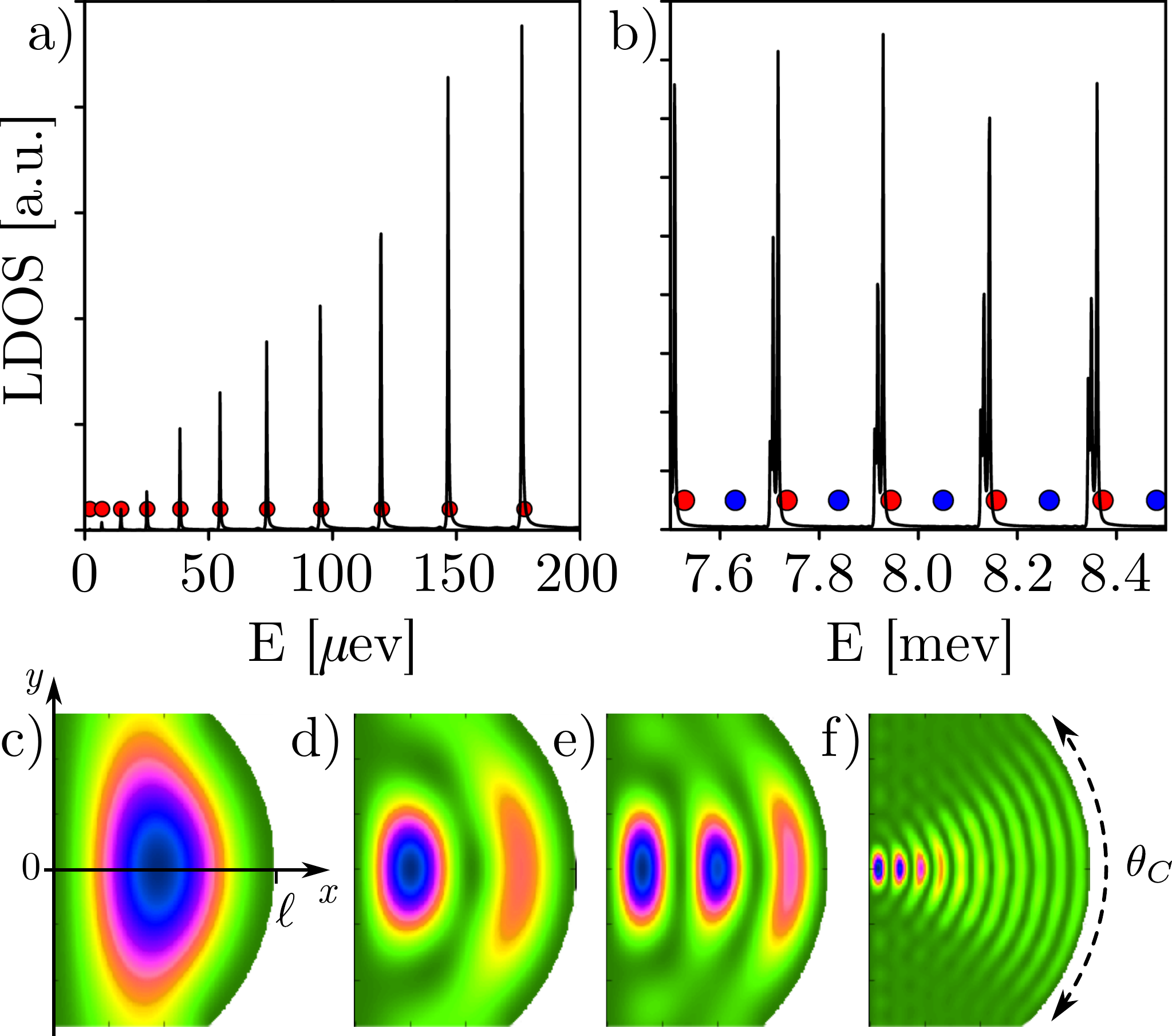}
  \caption{(a) LDOS calculated via Kwant near $r=0$ as a function of energy for  
ranges around (a) low energy, and (b) energy near $\varepsilon_F = 8$~meV. 
In (a) the peaks match the $n=1$ Bessel mode energies indicated by the red circles. 
In (b) the red (blue) dots are the odd (even) $l$ Bessel mode energies.
The data corresponds to a grid step size of 2~nm, which is close to numerical convergence.
The energy spacing between the LDOS peaks match $\delta_{cav}$ between odd $l$ modes (red dots).
The agreement with the odd Bessel modes improve as the step size is reduced, but a small discrepancy 
can be expect due to the real energy shift introduced by the self-energy of the leads.
(c)-(f) Full LDOS map for first peaks in panel (a), with energies $\sim 2, 7, 15$, and $180\mueV$.
}
  \label{fig:KwantLDOSvsLowE}
\end{figure}
%

We compare the Bessel function approximation with a numerically calculated 
LDOS implemented using the Kwant code \cite{Supp::kwant}. 
Figure \ref{fig:KwantLDOSvsLowE} shows a remarkably good agreement for both low energies 
and energies close the $\varepsilon_F$, corresponding to panels (a) and (b). 
Note that the even $l$ states (blue dots) do not contribute to the LDOS near $r=0$ as expected 
from the discussion based on Bessel eigenmodes. 
Panels (c)-(f) show the full LDOS map on the cavity for small energies, also in good agreement with 
the Bessel solutions shown in Fig.~\ref{fig:BesselDens}.

\section{Green's Functions and Equations of Motion}
\label{sec:EOM}

Our approach combines the equations-of-motion (EOM) with the numerical renormalization group (NRG) method to 
find the linear response current in strongly interacting systems. The EOM method allows us to assess the  
``single-particle" interference processes for arbitrarily complicated geometries and cast them in terms of effective 
energy dependent hybridization functions. The NRG, on the other hand, provides an robust approach to tread 
strongly correlated many-body systems and is amenable for including non-trivial geometric effects beyond the
wide band limit.

Before presenting the details of the calculation of the current, let us address the hybridization function of the 
experimental system of interest  and discuss some of the implications of our findings.

Let us begin by writing the Green's functions in the Zubarev notation, namely,
\begin{equation}
G_{A,B}(\omega) \equiv \langle\langle A; B \rangle\rangle_\omega\, ,
\end{equation}
with the corresponding equations of motion (EOMs)
\begin{align}
\omega  \langle\langle A; B \rangle\rangle = &\, 
\langle\{ A; B \}\rangle + \langle\langle [A,H] ; B \rangle\rangle
\nonumber\\
= &\, 
\langle\{ A; B \}\rangle - \langle\langle A ; [B,H] \rangle\rangle 
\end{align}
that have the same form for the retarded, advanced, and time-ordered  Green's 
functions (GFs). These GFs are computed for all combinations of creation and
annihilation operators in our model system. (The later correspond to 
$d_{\sigma}, 
c_{\alpha {\bf k} \sigma}$, and 
$a_{j\sigma}$ that are defined in the main text.)
In what follows, we shall omit the spin label $\sigma$, and indicate 
the type of Green's function only when necessary.

Using these results, one readily obtains a set of coupled Green's functions for 
our model Hamiltonian, defined in paper. These read
\begin{align}
\label{eq:Gjd}
(\omega - \ve_i) G_{id}(\omega) = &\,
\sum_{\bf k} V_{j{\bf k}} G_{R{\bf k}, d} (\omega) + V^*_{id} G^{}_{dd}(\omega),
\\
\label{eq:Gkd}
(\omega - \ve_{\alpha{\bf k}}) G_{\alpha{\bf k},d}(\omega) = &\,
 V_{\alpha d}^* G^{}_{d d} (\omega) + \delta_{\alpha R} \sum_{i}V^*_{iR} 
G_{id}(\omega),
\\
\label{eq:Gij}
(\omega - \ve_i) G_{i,j}(\omega) = &\,
 \delta_{ij} + V_{i d}^* G_{d i} (\omega) +  \sum_{{\bf k}}V_{iR} G_{R{\bf 
k},j}(\omega),
\\
\label{eq:Gkj}
(\omega - \ve_{\alpha{\bf k}}) G_{\alpha{\bf k},j}(\omega) = &\,
V_{d \alpha }^* G_{d j} (\omega) + \delta_{\alpha R} \sum_{i}V^*_{iR} 
G_{ij}(\omega).
\end{align}
We use the indices $i$ and $j$ to label cavity modes and $d$ to denote the quantum dot 
level. In the main text,  we use the standard shorthand notation $G_d \equiv G_{dd}$ for the 
quantum dot Green's function.

Using the expressions above, we can ``close" the EOMs (for $U=0$) and write the retarded quantum dot 
Green's function for the fully connected system in the absence of electron-electron interactions as 
\begin{equation}
G^{(0),r}_d(\omega) =\frac{1}{\omega - \ve_d - \Sigma^{(0),r}_d(\omega)}, 
\end{equation}
where the expression for $\Sigma^{(0),r}_d(\omega)$ is given in the main text. We define the energy-dependent effective hybridization function $\Delta(\omega) \!\equiv\! - \mbox{Im } \Sigma^{(0),r}_d(\omega)$.

\section{Numerical calculations}

The exact $U\!=\!0$ analytical expression for $\Delta(\omega) \!\equiv\! - \mbox{Im } \Sigma^{(0),r}_d(\omega)$ is used as input in the $U\! \neq \!0$ NRG calculations to capture the Kondo regime. To this end, we make a slight simplification in the model and consider equal couplings between all cavity levels $i=1,N$ and the right reservoir. This amounts  into setting $V_{iR}=V_{cR}$ in Eqs.\ \ref{eq:Gkd}--\ref{eq:Gkj}. We will refer this approximation as the ``simplified model" hereafter. 

\begin{figure}[ht!]
\begin{center}
\includegraphics[width=0.49\columnwidth]{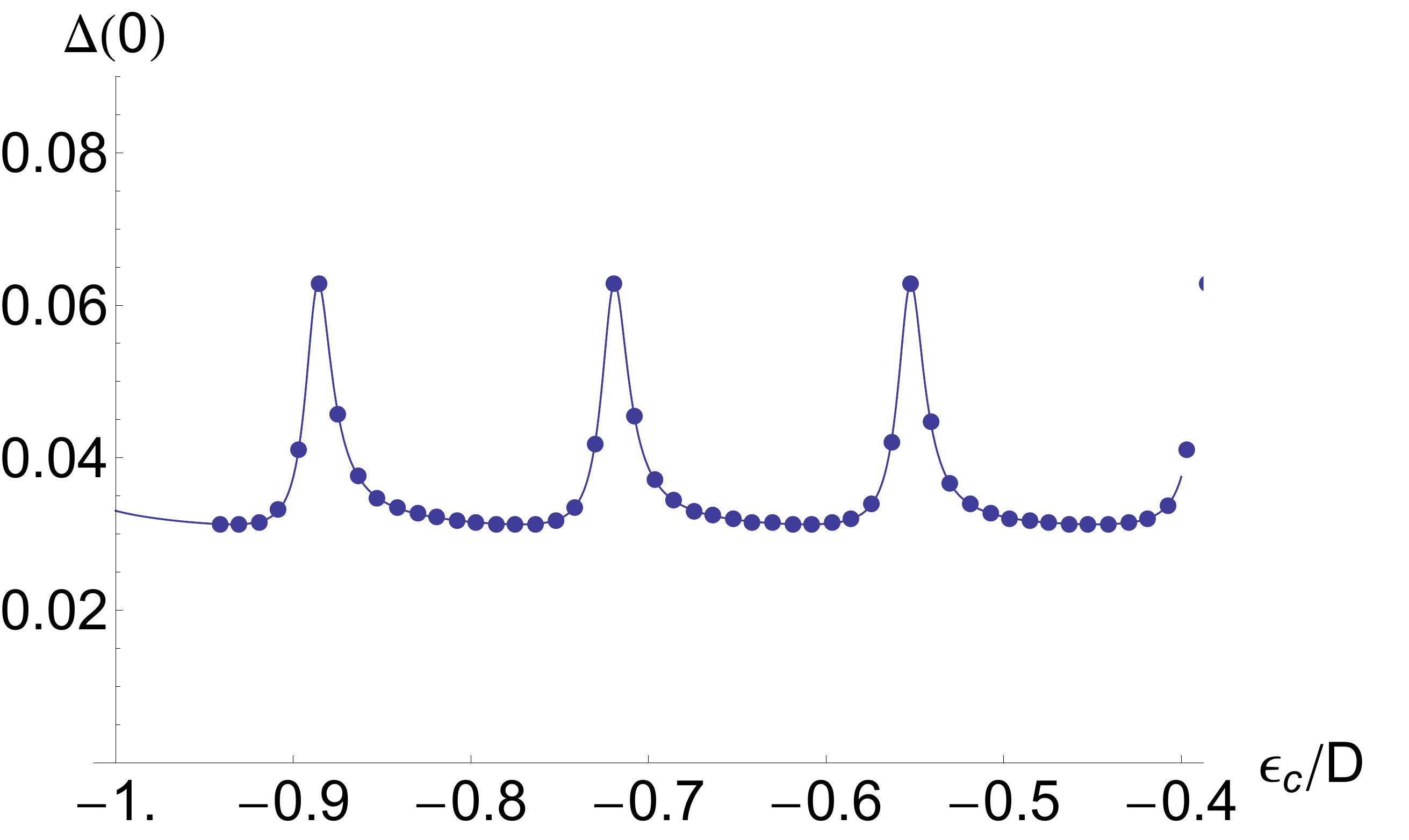}
\includegraphics[width=0.49\columnwidth]{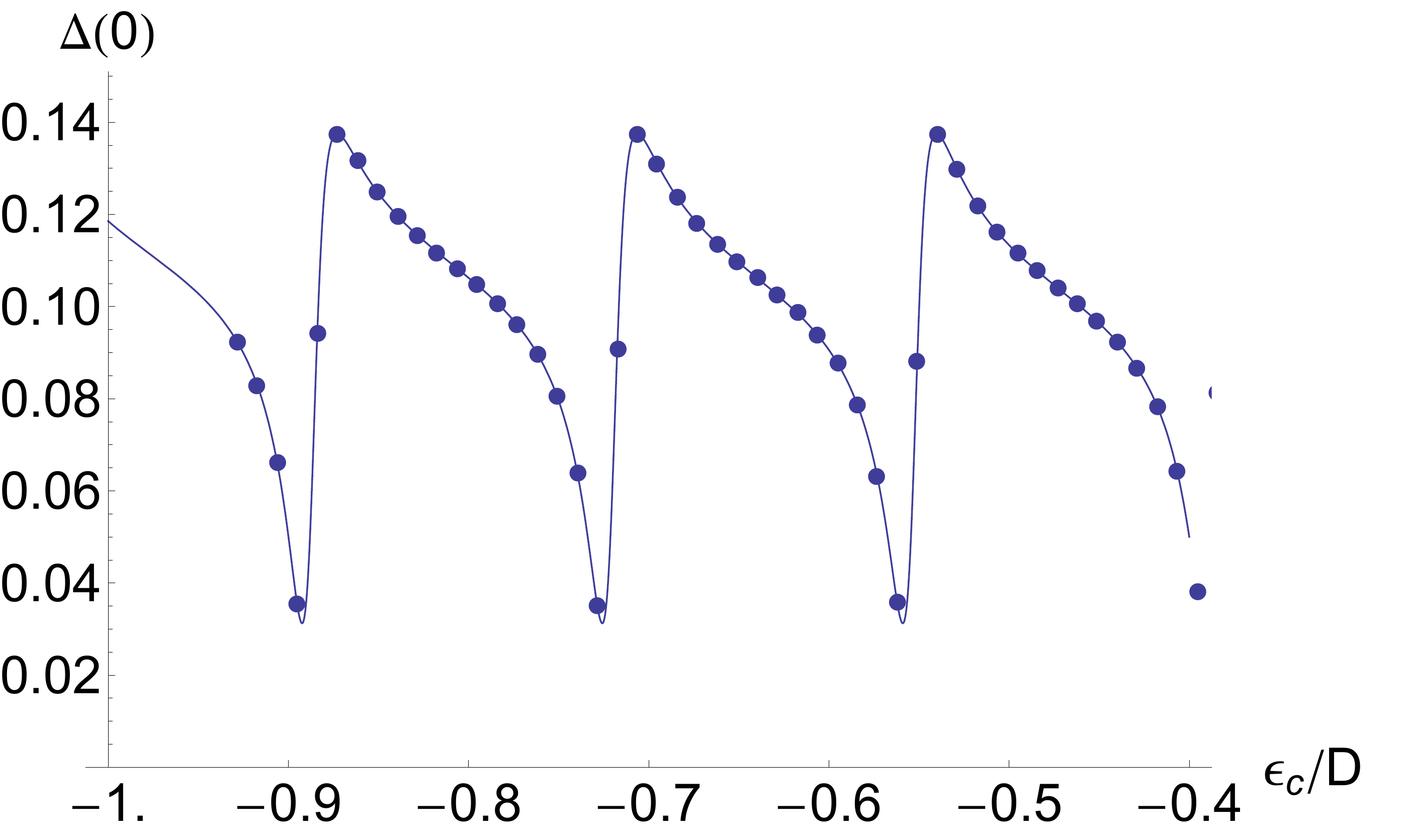}
\caption{(color online) Thin line: Effective hybridization function at the  
Fermi energy $\Delta(\omega\!=\!0)$ for weak ($\Omega=0.01D$, left) and strong 
($\Omega=0.15D$, right) coupling of the QD to the cavity, as function of the cavity gate-voltage 
$\epsilon_c/D$. Filled circles indicate the $\epsilon_c/D$ values points used in Fig. 2 of the main text. }  
\label{fig:Delta0_points}
\end{center}
\end{figure}

A key ingredient influencing the interacting do spectral function if the value of the hybridization function at the Fermi energy $\Delta(\omega=0)$ ($\omega\!=\!0$ is the Fermi level). Illustrative examples of $\Delta(0)$
vs  $\epsilon_c$, are shown in Fig.~\ref{fig:Delta0_points} for both weak ($\Omega=0.01D$) and strong-coupling ($\Omega=0.15D$) regimes. The other parameters used are those mentioned in the main paper, namely $\Gamma _{dL}\!=\!\Gamma_{dR}\!=\!0.0625 D$, $\Gamma _{cR} \!=\! 0.6 D$ and $\delta_{cav}\!=\!0.16 D$.

The drastically different dependence on $\epsilon_c$ in both cases is also reflected in contrasting 
$\omega$ dependence at fixed cavity parameters (not shown), which strongly affects the
effective spin fluctuations that set in once interactions are considered.  As we will show below,
this behavior has important consequences for the zero-bias 
conductance of the system, among other observables.

From the geometry of the device and the size of the cavity, it is natural to expect the cavity-reservoir coupling  to be much larger than the dot-reservoir coupling, such that $\Gamma_{cR} \gg \Gamma_{dR}$. Surprisingly, as a result of interference effects in the structure of $\Delta(\omega)$, such relative large cavity-reservoir couplings  translate into small widths in the peaks of $\Delta(0)$ in the weak cavity-dot coupling regime . In fact, the calculated widths of the peaks in Fig.\ \ref{fig:Delta0_points}-a ($\Omega=0.01D$) are $\Delta \epsilon_c \sim 0.022 D \ll \Gamma_{cR}$.

One can show that, in the non-interacting expression, the widths of the peaks in $\Delta(0)$  roughly translate into the width of the conductance peaks through the device in the weakly cavity-dot coupling regime. These were dubbed ``$\Gamma_{\rm cav}$" in Ref. \onlinecite{Supp::Klaus2015}. Using the experimental estimate of $U \sim 700 \mu$eV and taking $D=2U$, the widths in Fig.\ \ref{fig:Delta0_points}-a are $\approx 31 \mu$ eV, which is comparable to the experimental value for the conductance peak broadening  $``\Gamma _{\rm cav}"\! \sim \! 40 \mu$eV in Ref.\ \onlinecite{Supp::Klaus2015}. As we show in the text, the widths of the \textit{interacting} conductance peaks in the weak dot-cavity coupling regime are of the same order $\sim 56 \mu$eV.

\subsection{Details of the NRG calculations}

The NRG calculations were carried out using an effective single-site Anderson model for a symmetric impurity ($\ve_d=-U/2$) with an hybridization function given by $\Delta(\omega)$. The discretization of the effective band was carried out as discussed in Refs.\onlinecite{Supp::Gonzalez-Buxton98,Supp::Silva2008,Supp::Silva2006} using a discretization parameter $\Lambda=2.5$ and z-trick averaging ($N_z=5$). In the calculations, we explored charge and $SU(2)$ spin symmetries and up to 1000 $Q,S$ states were retained at each NRG iteration. 

The spectral density data shown in the paper were obtained using the DM-NRG method.\cite{Hofstetter:1508:2000} Additional runs using the CFS approach\cite{Peters:Phys.Rev.B:245114:2006, Weichselbaum:Phys.Rev.Lett.:99:076402:2007} were also performed to check convergence of the results.

\begin{figure}[ht!]
\begin{center}
\includegraphics[width=0.49\columnwidth]{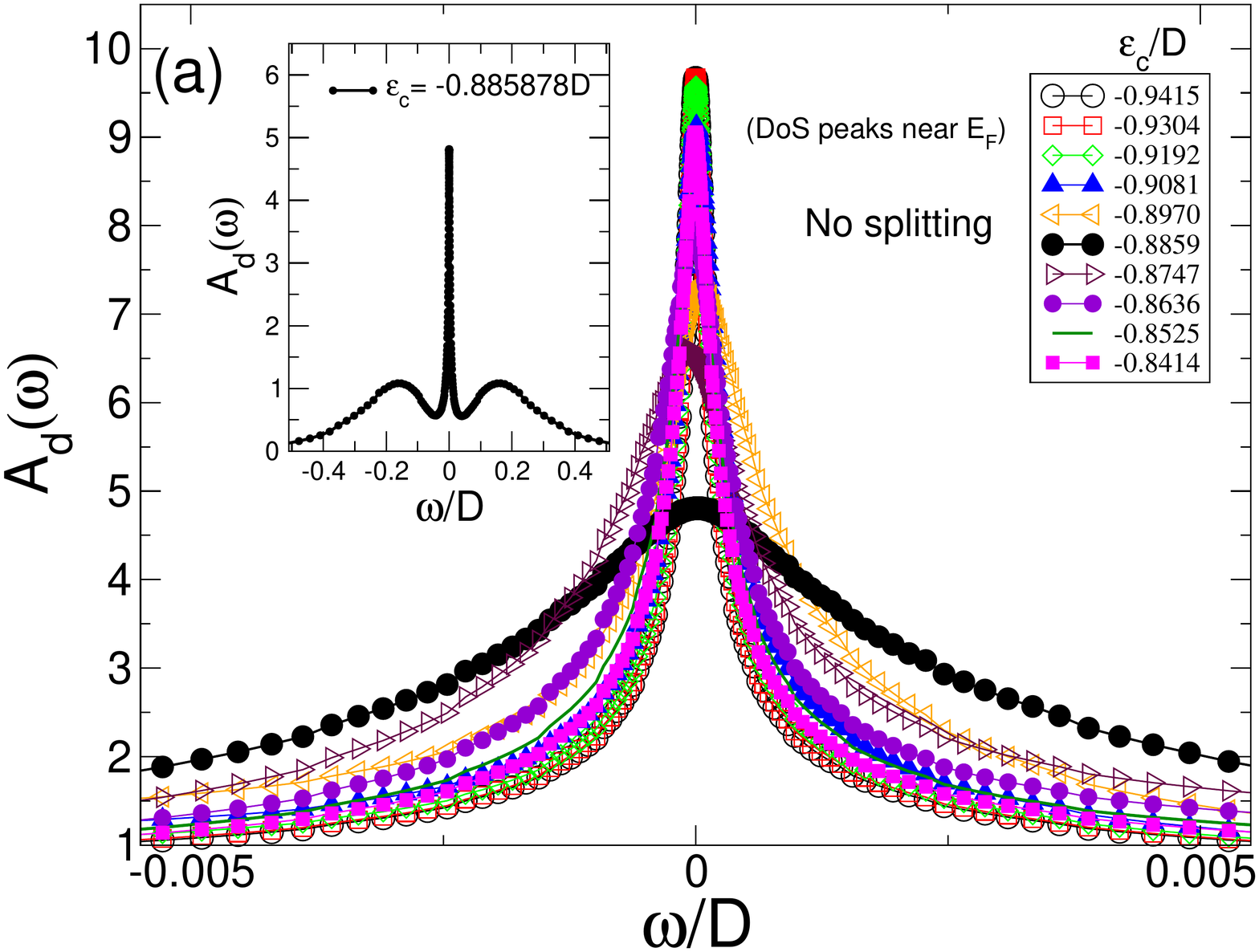}
\includegraphics[width=0.49\columnwidth]{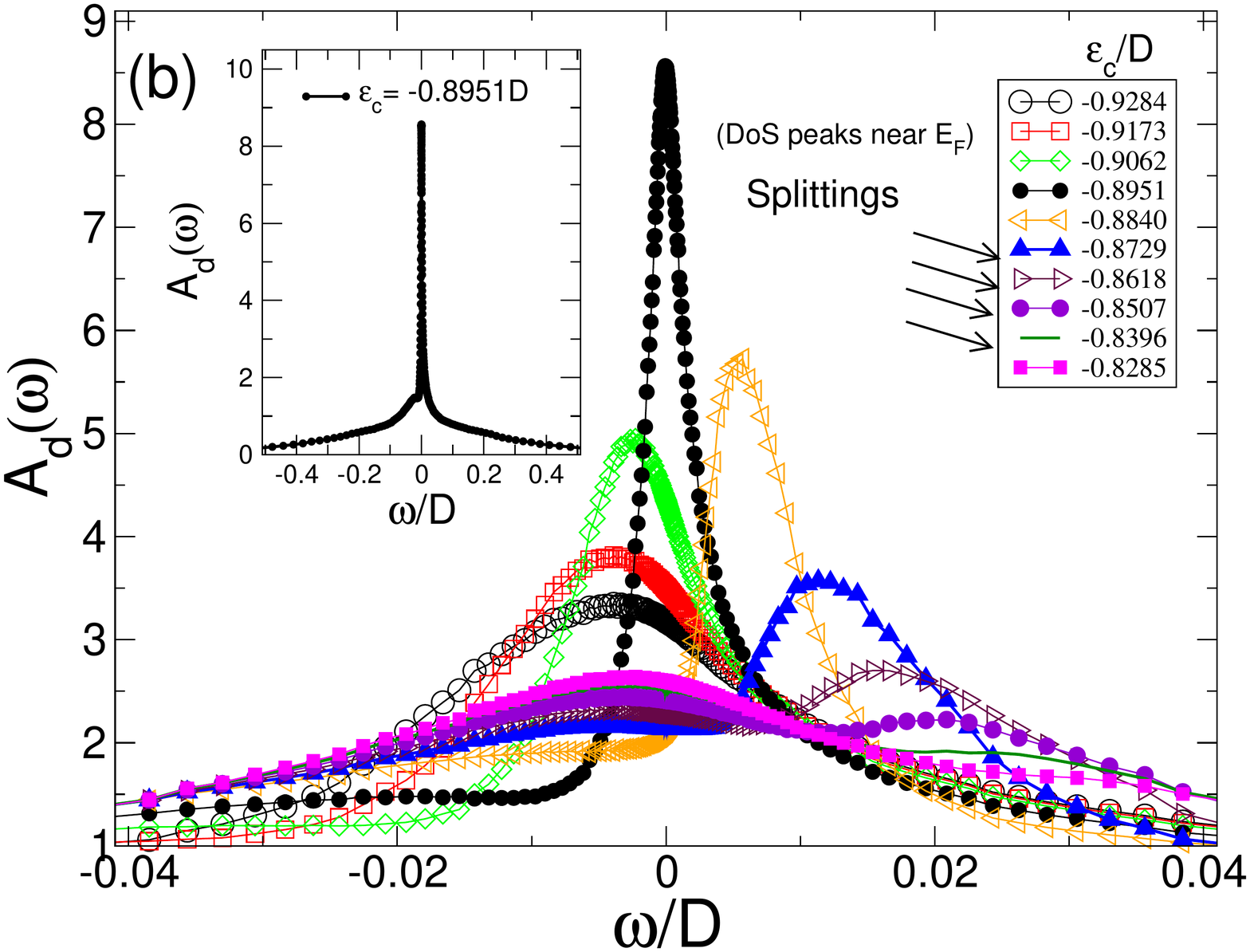}
\caption{(color online) NRG spectral functions for (a) $\Omega\!=\!0.01D$ and (b) $\Omega\!=\!0.15D$ and different values of $\epsilon_c$ (same data is shown as a contour plot in Fig. 2 of the main text). Very narrow peaks near $\omega\!=\!0$ appear as a result of peak splitting due to the cavity-originated resonances in $\Delta(0)$.  The insets show a typical data for $\epsilon_c$ values at the resonances of $\Delta(0)$ (see Figs. \ref{fig:Delta0_points}). Notice the broader resonances for $\Omega\!=\!0.15D$ [panel (b)] indicating larger Kondo temperatures.  }  
\label{fig:SpecDens_Cavity}
\end{center}
\end{figure}

Examples for the results (data in Fig. 2 of the main paper) are presented in Fig.\ \ref{fig:SpecDens_Cavity}. Notice the formation of the Kondo resonance in the insets, with a broader peak for for $\Omega\!=\!0.15D$ indicating a larger $T_K$, as discussed in the main text.

\section{Calculation of the current through the system}

\subsection{Extension of the Meir-Wingreen formalism}

In general, the current flowing from the  $\alpha$ contact can be written as 
\begin{equation}
J_\alpha = -e \left\langle \frac{d}{dt} N_\alpha \right\rangle
\end{equation}
where $N_\alpha = \sum_{{\bf k} \sigma}  c_{\alpha{\bf k}\sigma}^\dagger 
c_{\alpha{\bf k}\sigma}^{}$ counts the number of electrons at lead $\alpha$. 
Let us start with the right lead ($\alpha = R$). Using the Heisenberg  picture, 
where $i\hbar \dot{c}_{R{\bf k}\sigma} = [c_{R{\bf k}\sigma}, H ]$, one obtains
\begin{equation}
\label{eq:JR_time}
J_R (t) = -e \sum_{{\bf k}\sigma} \Big[  \sum_i V_{iR}^{}  G^<_{R{\bf k}\sigma, 
i\sigma} (t,t) + 
V_{dR}^{} G^<_{R{\bf k}\sigma, d\sigma}(t,t) - {\rm H.c.} \Big],
\end{equation}
where the following Green's functions were introduced,
\begin{align}
G^<_{R{\bf k}\sigma, i\sigma} (t,t') = &
\frac{i}{\hbar} \langle a^\dagger_{i\sigma}(t') c_{R{\bf k}\sigma}^{}(t) 
\rangle,\\
G^<_{R{\bf k}\sigma, d\sigma} (t,t') = &
\frac{i}{\hbar} \langle c^\dagger_{d\sigma}(t') c_{R{\bf k}\sigma}^{}(t) \rangle.
\end{align}
The current $J_R$ is real, since \cite{Haug96} $G_{ab}^<(t,t)=-[G_{ba}^<(t,t)]^*$. 

We are interested in the stationary regime, where $J_R$ does not depend on time. 
Thus, it is convenient to write Eq.~\eqref{eq:JR_time} in the frequency 
representation, 
\begin{equation}
\label{eq:JRexpression0}
 J_R = 2e\, {\rm Re}\Bigg\{ \sum_{{\bf k}\sigma} \int \!\frac{d\omega}{2\pi \hbar} 
\Bigg[ \sum_i V_{iR} ^{}G^<_{R{\bf k}\sigma,i \sigma} (\omega) + \\ 
  V_{dR}^{} G^<_{R{\bf k}\sigma,d\sigma} (\omega) \Bigg]\Bigg\}.
\end{equation}
The above equation is the generalization of the two-terminal Meir-Wingreen 
formula \cite{Supp::MeirWingreen92} for our model system, where the right lead 
($\alpha = R$) is coupled to both the dot and the cavity; see 
Fig.~1 of the paper.

In contrast, the left lead ($\alpha = L$) is only coupled to the dot. 
Consequently, the current $J_L$ is given by the standard expression 
\begin{equation}
\label{eq:JLexpression0}
J_L \!=\!  2e\, {\rm Re}\left\{ \sum_{{\bf k}\sigma} \int \!\frac{d\omega}{2\pi 
\hbar} \left[ V_{dL}^{} G^<_{L{\bf k}\sigma,d\sigma} (\omega) \right]\right\}.  
\end{equation}


Next, we use the method of  equations of motion (EOM) and the Langreth rules \cite{Supp::Langreth,Langreth:Book::1976}
to express the Green's function $G^<_{i\sigma,R{\bf k}\sigma} (\omega)$ in a 
convenient form. 

Using the results of Section \ref{sec:EOM}, the contact Green's functions $G_{R{\bf k}, d}$ and $G_{R{\bf k}, j}$ that 
appear 
in Eq.~\eqref{eq:JRexpression0} can be expressed as
\begin{equation}
\label{eq:Gkd_better}
G_{R{\bf k}, d}(\omega) 
= \frac{V^*_{dR}}{\omega - \ve_{R{\bf k}} }   G_{dd}(\omega)
+ \frac{1}{\omega - \ve_{R{\bf k}} } \sum_i V^*_{iR} G_{id}(\omega)
= g_{R{\bf k}}(\omega) V^*_{dR} G_{dd}(\omega)
+ g_{R{\bf k}}(\omega) \sum_i V^*_{iR} G_{id}(\omega)
\end{equation}
and
\begin{equation}
\label{eq:Gkj_better}
G_{R{\bf k}, j}(\omega) 
= \frac{V^*_{dR}}{\omega - \ve_{R{\bf k}} }  G_{dj}(\omega)
    +   \frac{1}{\omega - \ve_{R{\bf k}} } \sum_i V^*_{iR} G_{ij}(\omega)
= g_{R{\bf k}}(\omega) V^*_{dR} G_{dj}(\omega)
+     g_{R{\bf k}}(\omega)  \sum_i V^*_{iR} G_{ij}(\omega),
\end{equation}
where $g_{R{\bf k}}(\omega)$ is the free Green's function at the terminal $R$.

Recall that in the simple two-terminal case one has to deal only 
with $G_{R{\bf k}, d} = g_{R{\bf k}}^{} V^*_{dR} G_{dd}^{}$. This means that 
the problem  is reduced to the calculation of $G_{dd}$, see Section 
\ref{app:MW-like}.  Our goal here is similar: we want to eliminate all hybrid 
(or contact) Green's function and express the current in terms of $G_{dd}$ only. This is always possible, as long as interactions are local and restricted to the QD.

Let us now solve for $G_{jd}$. By inserting Eq.~\eqref{eq:Gkd_better} into 
\eqref{eq:Gjd} we write 
\begin{equation}
(\omega -\ve_j) G_{jd}(\omega) \!=\! V^*_{jd} G_{dd}(\omega)+
\sum_{\bf k} \frac{V_{jR} V^*_{dR}}{\omega - \ve_{R{\bf k}}}  G_{dd}(\omega) 
+ \sum_{{\bf k},i} V_{jR} \frac{1}{\omega - \ve_{R{\bf k}}} V^*_{iR}
G_{id}(\omega).
\end{equation}
Hence
\begin{equation}
\sum_i \left[ (\omega-\ve_i) \delta_{ij} - \sum_{\bf k} V_{jR} \frac{1}{\omega 
- \ve_{R{\bf k}}} V^*_{iR}\right] \!G_{id}(\omega) =\\ \left[ V_{jd} + \sum_{\bf 
k} V_{jR} \frac{1}{\omega - \ve_{R{\bf k}}} V^*_{dR}\right] G_{dd}(\omega)
\end{equation}

Next, let us solve for $G_{ij}$. By inserting Eq.~\eqref{eq:Gkj_better} 
into \eqref{eq:Gij} we get 
\begin{equation}
(\omega -\ve_i) G_{ij}(\omega) = \delta_{ij} +  \sum_{\bf k} V_{iR} 
\frac{1}{\omega - \ve_{R{\bf k}}} V^*_{dR} G_{dj}(\omega) +
\sum_{{\bf k},l} V_{iR} \frac{1}{\omega - \ve_{R{\bf k}}} V^*_{lR} 
G_{lj}(\omega) + 
V^*_{id} G_{dj}(\omega)\,.
\end{equation}
Hence,
\begin{equation}
\sum_l \left[ (\omega-\ve_l) \delta_{lj} - \sum_{\bf k} V_{iR} \frac{1}{\omega 
- \ve_{R{\bf k}}} V^*_{lR}\right] \!G_{lj}(\omega) = \delta_{ij} 
+
\left[ V_{id} + \sum_{\bf k} V_{iR} \frac{1}{\omega -  \ve_{R{\bf k}}} 
V^*_{dR}\right] \!G_{dj}(\omega) .\,\,\,
\end{equation}

Before we proceed, let us simplify the notation by introducing the 
resonance self-energies
\begin{equation}
\label{eq:ResonanceSelfEnergies}
\Sigma_{\nu\nu'}^\alpha(\omega) = \sum_{\bf k} 
V_{\nu \alpha}^{} \frac{1}{\omega - \ve_{\alpha{\bf k}}} V^*_{\nu'\alpha}
\end{equation}
where $\nu=i,d$. Let us also define
\begin{equation}
\label{eq:G0ij}
\sum_l \left[ (\omega-\ve_l) \delta_{li} - 
\Sigma_{il}^R(\omega) \right] \!G_{lj}^{(0)}(\omega) = 
\delta_{ij}.
\end{equation}
Collecting the results, we obtain
\begin{equation}
\label{eq:G_id}
G_{id}(\omega) = \sum_j G^{(0)}_{ij}(\omega) 
\left[ V_{jd} + \Sigma_{jd}^R(\omega)\right] \!G_{dd}(\omega)
\end{equation}

and
%
\begin{equation}
\label{eq:G_ij}
G_{ij}(\omega) = \, G^{(0)}_{ij}(\omega)  + \sum_l G^{(0)}_{il}(\omega) 
\left[ V_{ld} + \Sigma_{ld}^R (\omega) \right] \!G_{dj}(\omega).
\end{equation}
%
%


Note that the integrand in Eq.~\eqref{eq:JRexpression0} contains the Green's functions
$G^<_{i\sigma,R{\bf k}\sigma} (\omega)$ and $G^<_{d\sigma,R{\bf k}\sigma}(\omega)$.
Using the  Langreth rules \cite{Supp::Langreth}
and the  Eqs.~\eqref{eq:Gkd_better}  and \eqref{eq:Gkj_better}  we 
write
\begin{equation}
G^<_{d,R{\bf k}} (\omega)= G_{dd}^r(\omega) V^{}_{dR} g_{R{\bf 
k}}^<(\omega)+ G_{dd}^<(\omega) V^{}_{dR} g_{R{\bf k}}^a(\omega) 
 +\sum_i G_{id}^r(\omega)  V^{}_{iR} g_{R{\bf k}}^<(\omega) + \sum_i 
G_{id}^<(\omega)  V^{}_{iR} g_{R{\bf k}}^a(\omega)
\end{equation}
and
\begin{equation}
G_{j,R{\bf k}}^<(\omega)= G_{jd}^r(\omega)  V^{}_{dR} g_{R{\bf 
k}}^<(\omega) + G_{jd}^<(\omega)  V^{}_{dR} g_{R{\bf k}}^a(\omega)  + \sum_i \left[G_{ji}^r(\omega)  V^{}_{iR}   g_{R{\bf 
k}}^<(\omega) +
G_{ji}^<(\omega) V^{}_{iR}   g_{R{\bf k}}^a(\omega) \right]
\end{equation}
where the free Green's functions are given by
\begin{align}
g_{\alpha{\bf k}}^{r(a)}(\omega) = \frac{1}{\omega -  \ve_{\alpha{\bf k}} \pm 
i0} 
= \mbox{PV} \frac{1}{\omega - \ve_{\alpha{\bf k}}}  \mp i \pi \delta(\omega - 
\ve_{\alpha{\bf k}}) 
\end{align}
and $g_{\alpha{\bf k}}^{<}(\omega) =  2\pi i \delta(\omega  - \ve_{\alpha{\bf 
k}}) f_\alpha(\omega)$.

In the wide band limit, we can evaluate the self-energies as
\begin{equation}
\label{eq:ResonanceSelfEnergies-flat}
\Sigma_{\nu\nu'}^{\alpha,r/a}(\omega) = \mp i\pi  V_{\nu\alpha} \rho_\alpha V^*_{\nu^\prime\alpha}
\end{equation}
where $\rho_\alpha$ is density of states of the reservoir $\alpha=R,L$.  
From this expression we can define  $\Sigma_{\nu\nu}^{\alpha,r}(\omega)=
- i\pi  V_{\nu\alpha} \rho_\alpha V^*_{\nu\alpha}\equiv -i\Gamma_{\nu\alpha}/2$,
and assuming the couplings real, 
$\Sigma_{id}^{\alpha,r}(\omega)=-i\pi  V_{i\alpha} \rho_\alpha V^*_{d\alpha}=
-i\sqrt{\Gamma_{iR}\Gamma_{dR}}/2$. Notice that this definition of $\Gamma_{\nu\alpha}$ (more frequently used in transport works) carries an extra factor of 2 as compared to the definition commonly used by the strongly-correlated systems community (``$\Gamma= \pi \rho |V|^2$" ).

We now introduce the new self-energies
\begin{eqnarray}
\label{eq:TildeSelfEnergies1}
\widetilde{\Sigma}_{jd}^{R,a}(\omega) = &\, V_{jd} + i \sqrt{\Gamma_{jR}\Gamma_{dR}}/2,
\\
\label{eq:TildeSelfEnergies2}
\widetilde{\Sigma}_{jd}^{R,r}(\omega) = &\, V_{jd} - i\sqrt{\Gamma_{jR}\Gamma_{dR}}/2,
\end{eqnarray}
and
\begin{eqnarray}
\widetilde{\Sigma}_{jd}^{R,<}(\omega) = &\, + i f_R(\omega) \sqrt{\Gamma_{jR}\Gamma_{dR}}.
\label{eq:TildeSelfEnergies3}
\end{eqnarray}

Using the Langreth rules\cite{Supp::Langreth} we are able to express $G_{id}$ and $G_{ij}$, given by 
Eqs.~\eqref{eq:G_ij} and \eqref{eq:G_id},  in terms of free Green's functions 
(that we know analytically) and of $G_{dd}$. 
Combining Eqs.~\eqref{eq:G_id} and \eqref{eq:G_ij}  with \eqref{eq:TildeSelfEnergies1}-
\eqref{eq:TildeSelfEnergies3} we can write
\begin{align}
\label{eq:Gidra}
G_{id}^{r(a)}(\omega) =&\, \sum_j G^{(0),r(a)}_{ij}(\omega) \,
\widetilde{\Sigma}_{jd}^{R,r(a)}(\omega)\, G_{dd}^{r(a)}(\omega),
\end{align}
%
\begin{eqnarray}
G_{id}^{<}(\omega) &=&\, \sum_j \left[ 
G^{(0),r}_{ij}(\omega) \widetilde{\Sigma}_{jd}^{R,r}(\omega)G_{dd}^{<}(\omega) +
G^{(0),r}_{ij}(\omega) \widetilde{\Sigma}_{jd}^{R,<}(\omega)G_{dd}^{a}(\omega) +
G^{(0),<}_{ij}(\omega) \widetilde{\Sigma}_{jd}^{R,a}(\omega)G_{dd}^{a}(\omega)\right], 
\end{eqnarray}
\begin{align}
\label{eq:Gdi_lesser}
G_{di}^{<}(\omega) = \sum_j  \left[
G_{dd}^{r}(\omega) \widetilde{\Sigma}_{dj}^{R,r}(\omega) G^{(0),<}_{ji}(\omega) +
G_{dd}^{r}(\omega) \widetilde{\Sigma}_{dj}^{R,<}(\omega) G^{(0),a}_{ji}(\omega) +
G_{dd}^{<}(\omega) \widetilde{\Sigma}_{dj}^{R,a}(\omega) G^{(0),a}_{ji}(\omega) 
\right],
\end{align}
and
\begin{eqnarray}
 G_{ij}^{<}(\omega)&=&G_{ij}^{(0),<}(\omega)+\sum_{ll^\prime} 
   \left[  G_{il}^{(0),r}(\omega) 
\tilde\Sigma_{ld}^{R,r}(\omega)G_{dd}^{r}(\omega)\tilde 
      \Sigma_{dl^\prime}^{R,r}(\omega) G_{l^\prime j}^{(0),<}(\omega) + 
G_{il}^{(0),r}(\omega)\tilde\Sigma_{ld}^{R,r}(\omega)G_{dd}^{r}(\omega)\tilde 
      \Sigma_{dl^\prime}^{R,<}(\omega) G_{l^\prime j}^{(0),a}(\omega) 
\right.\nonumber\\    
  && + G_{il}^{(0),r}(\omega)\tilde\Sigma_{ld}^{R,r}(\omega)G_{dd}^{<}(\omega)\tilde 
      \Sigma_{dl^\prime}^{R,a}(\omega) G_{l^\prime j}^{(0),a}(\omega) + 
G_{il}^{(0),r}(\omega)\tilde\Sigma_{ld}^{R,<}(\omega)G_{dd}^{a}(\omega)\tilde 
    \Sigma_{dl^\prime}^{R,a}(\omega) G_{l^\prime j}^{(0),a}(\omega) 
\nonumber\\    
 &&\left.+ 
G_{il}^{(0),<}(\omega)\tilde\Sigma_{ld}^{R,a}(\omega)G_{dd}^{a}(\omega)\tilde 
\Sigma_{dl^\prime}^{R,a}(\omega) G_{l^\prime j}^{(0),a}(\omega) \right].
\end{eqnarray}

We are now ready to return to Eq.~\eqref{eq:JRexpression0}  and calculate the 
current $J_R \equiv J_R^{(1)} + J_R^{(2)}$, with
\begin{eqnarray}
J_R^{(1)} = 2e\sum_{{\bf k} \sigma} \sum_i{\rm Re} \int 
\!\frac{d\omega}{2\pi \hbar} \Big\{  G_{id}^r(\omega)  V^{}_{dR} 2\pi i 
\delta(\omega-\ve_{\bf k}) f_R(\omega) V_{iR}^*+ G_{id}^<(\omega) V^{}_{dR} i\pi   
\delta(\omega-\ve_{\bf k} ) V_{iR}^* 
\nonumber\\ 
+  \sum_j\left[G_{ij}^r(\omega)  V^{}_{jR}  2\pi i \delta(\omega-\ve_{\bf 
k}) f_R(\omega)  V_{iR}^*+ G_{ij}^<(\omega) V^{}_{jR}i\pi\delta(\omega-\ve_{\bf k} 
) V_{iR}^*\right]\Big\}
\end{eqnarray}
and
\begin{eqnarray}
J_R^{(2)}=2e\sum_{{\bf k} \sigma} {\rm Re} \int \!\frac{d\omega}{2\pi 
\hbar} \Big\{G_{dd}^r(\omega)  V^{}_{dR} 2\pi i \delta(\omega-\ve_{\bf k}) 
f_R(\omega) V_{dR}^*+ G_{dd}^<(\omega) V^{}_{dR} i\pi \delta(\omega-\ve_{\bf k} ) 
V_{dR}^* \nonumber\\ 
+\sum_i \left[G_{di}^r(\omega)  V^{}_{iR}  2\pi i \delta(\omega-\ve_{\bf k}) 
f_R(\omega)  V_{dR}^*+ G_{di}^<(\omega) V^{}_{iR}  i\pi \delta(\omega-\ve_{\bf k} ) 
V_{dR}^*\right]\Big\},
\end{eqnarray}
%
where have used the wide flat band approximation to get rid of the Cauchy 
principal value contribution.

We now convert the summations over ${\bf k}$ into energy integrations, namely 
\begin{eqnarray}
\sum_{\bf k} \Big[ \cdots \Big] = \int d\ve_{\bf k} \rho_\alpha (\ve_{\bf k}) 
\Big[ \cdots \Big].
\end{eqnarray}

For notational simplicity, let us assume  that all coupling matrix elements $V$ 
are real to write
%
\begin{align}
J_R^{(1)} = & 
\frac{e}{\hbar}\sum_{i \sigma} {\rm Re} \!\int \!\frac{d\omega}{2\pi} 
          \Big\{    i  \sqrt{\Gamma_{dR}\Gamma_{iR}} \big[2 f_R(\omega) 
G_{id}^r(\omega)+
          G_{jd}^<(\omega) \big] +
  \sum_j  i \sqrt{\Gamma_{jR}\Gamma_{iR}} \left[2 f_R(\omega) G_{ij}^r(\omega)   +
                    G_{ij}^<(\omega) \right]
        \Big\}
\nonumber\\ =&\,  
\frac{e}{\hbar}\sum_{i \sigma}  \int \!\frac{d\omega}{2\pi} 
          \Big\{    \sqrt{\Gamma_{dR}\Gamma_{iR}} \left[2 f_R(\omega) {\rm Re} [i 
G_{id}^r(\omega)]+
          {\rm Re}[i G_{jd}^<(\omega)] \right] +
  \sum_j  \sqrt{\Gamma_{jR}\Gamma_{iR}} \left[2 f_R(\omega) {\rm Re} [i 
G_{ij}^r(\omega)]   +
                   {\rm Re}[i G_{ij}^<(\omega)] \right]
        \Big\}
\end{align}
and
\begin{align}
J_R^{(2)} = & 
\frac{e}{\hbar}\sum_{\sigma} {\rm Re} \!\int \!\frac{d\omega}{2\pi} 
          \Big\{    i  \Gamma_{dR} \big[2 f_R(\omega) G_{dd}^r(\omega)+
          G_{dd}^<(\omega) \big] +
  \sum_j  i \sqrt{\Gamma_{dR}\Gamma_{jR}} \left[2 f_R(\omega) G_{dj}^r(\omega)   +
                    G_{dj}^<(\omega) \right]
        \Big\}
\nonumber\\ =&\,  
\frac{e}{\hbar}\sum_{\sigma}  \int \!\frac{d\omega}{2\pi} 
          \Big\{    \Gamma_{dR} \left[2 f_R(\omega) {\rm Re} [i G_{dd}^r(\omega)]+
          i G_{dd}^<(\omega) \right] +
  \sum_j  \sqrt{\Gamma_{dR}\Gamma_{jR}} \left[2 f_R(\omega) {\rm Re} [i 
G_{dj}^r(\omega)]   +
                   {\rm Re}[i G_{dj}^<(\omega)] \right]
        \Big\}.
\end{align}

We recall that $[G_{a,b}^<(\omega)]^* = - G_{b,a}^<(\omega)$. Therefore
$G_{dd}^<(\omega)$ is pure imaginary. For the non-diagonal terms we use Re$(ic)= 
i(c - c^*)/2$ to write
\begin{equation}
{\rm Re} [iG_{a,b}^<(\omega)] = \frac{i}{2}\left[ G_{a,b}^<(\omega) - 
[G_{a,b}^<(\omega)]^*\right]  =
\frac{i}{2}\left[ G_{a,b}^<(\omega) + G_{b,a}^<(\omega)\right].
\end{equation}
Therefore,
\begin{align}
J_R^{(1)} = 
\frac{i e}{h}\sum_{i \sigma}  \int \! d\omega \sqrt{\Gamma_{iR}}\, \Big\{ 
 & \sqrt{\Gamma_{dR}} \left[ f_R^{}(\omega)[G_{id}^r(\omega) 
-G_{id}^a(\omega)]+\frac{1}{2}\left( G_{id}^<(\omega) + G_{di}^<(\omega)\right) 
\right]\nonumber\\ &
 +\sum_j \sqrt{\Gamma_{jR}} \left[f_R(\omega) [G_{ij}^r(\omega) 
-G_{ij}^a(\omega)] + \frac{1}{2}\left( G_{ij}^<(\omega) + G_{ji}^<(\omega)\right) 
\right] \Big\}
\end{align}
and
\begin{align}
J_R^{(2)} = 
\frac{i e}{h}\sum_{\sigma}  \int \! d\omega \,\Big\{  & 
\Gamma_{dR} \left[ f_R^{}(\omega) [G_{dd}^r(\omega)-G_{dd}^a(\omega)]+  
G_{dd}^<(\omega) \right] \nonumber\\ &
 + \sum_j  \sqrt{\Gamma_{jR}\Gamma_{dR}}\left[ f_R(\omega) 
[G_{dj}^r(\omega)-G_{dj}^a(\omega)] + \frac{1}{2}\left(G_{di}^<(\omega) + 
G_{id}^<(\omega)\right) \right]\Big\}
\end{align}
and finally
\begin{align}
\label{eq:JR_intermediario}
J_R = \frac{i e}{h}\sum_{\sigma}  \int \! d\omega \, \Bigg\{  & 
\Gamma_{dR} \left[ f_R^{}(\omega) [G_{dd}^r(\omega)-G_{dd}^a(\omega)]+  
G_{dd}^<(\omega) \right]
\nonumber\\ &
+ \sqrt{\Gamma_{dR}} \,\sum_j  \sqrt{\Gamma_{jR}} 
 \left[ f_R(\omega) \left[G_{dj}^r(\omega)+G_{jd}^r(\omega)-G_{dj}^a(\omega)- 
G_{jd}^a(\omega)\right ]+ G_{dj}^<(\omega) + G_{jd}^<(\omega) 
\right] \nonumber\\ &
 + \sum_{ij}  \sqrt{\Gamma_{iR}\Gamma_{jR}} \left[ f_R(\omega) 
\left[G_{ij}^r(\omega)-G_{ij}^a(\omega)\right] +\frac{1}{2}\left( 
G_{ij}^<(\omega) + G_{ji}^<(\omega)\right) \right]\Bigg\}.
\end{align}
This lengthy expression reduces to the standard expression for the current 
found in Meir-Wingreen when one considers the simple case without cavity, 
that is, $\Gamma_{iR}=0$.

Using $G^a(\omega) = [G^r(\omega)]^\dagger$
one could simplify somewhat the second line of Eq.~\eqref{eq:JR_intermediario} 
to obtain 
%
%
\begin{eqnarray}
\label{eq:JR_intermediario1}
J_R &=& \frac{i e}{h}\sum_{\sigma}  \int \! d\omega \, \Bigg\{ \Gamma_{dR} 
\left[ f_R^{}(\omega) [G_{dd}^r(\omega)-G_{dd}^a(\omega)]+  G_{dd}^<(\omega) \right] 
\nonumber\\
 && +2\sqrt{\Gamma_{dR}} \,\sum_j  \sqrt{\Gamma_{jR}}\left[ f_R(\omega) 
\left[G_{jd}^r(\omega)-G_{jd}^a(\omega)\right ]+\frac{1}{2}\left( G_{jd}^<(\omega) + 
G_{dj}^<(\omega)\right) \right] \nonumber\\
 &&+ \sum_{ij}  \sqrt{\Gamma_{iR}\Gamma_{jR}}\left[f_R(\omega) 
\left[G_{ij}^r(\omega)-G_{ij}^a(\omega)\right] +\frac{1}{2}\left( G_{ij}^<(\omega) + 
G_{ji}^<(\omega)\right) \right]\Bigg\}.
\end{eqnarray}
%

The current from the left lead is simple because $\Gamma_{iL}=0$, so we 
have for $J_L$,
\begin{equation}
\label{eq:JL_intermediario2}
J_L = \frac{i e}{h}\sum_{\sigma}  \int  d\omega \, 
\Gamma_{dL}  \Big\{ f_L^{}(\omega) [G_{dd}^r(\omega)- G_{dd}^a(\omega)] 
 + G_{dd}^<(\omega) \Big\} .
\end{equation}

Notice that Eqs.\ (\ref{eq:JR_intermediario1}) and (\ref{eq:JL_intermediario2}) can be written in the matrix form used in Ref.\ \onlinecite{Supp::MeirWingreen92}: 

%
\begin{eqnarray}
\label{eq:JR_matrix}
J_{\ell=R,L} &=& \frac{i e}{h}\sum_{\sigma}  \int \! d\omega \, \Bigg( \mbox{ tr} \left\{ f_{\ell}^{}(\omega) \mathbf{\Gamma}^\ell \left[ \mathbf{G}^r-\mathbf{G}^a \right] \right\} + \mbox{ tr} \left\{ \mathbf{\Gamma}^\ell \mathbf{G}^< \right\} \Bigg).
\end{eqnarray}
%
where matrices $\mathbf{\Gamma}^{R,L}$ and the \textit{interacting} Green's functions $\mathbf{G}^{\alpha=r,a,<}$ are given by:

\begin{equation}
\mathbf{\Gamma}^R = \left(  
\begin{array}{cccc}
\Gamma_{dR} & \sqrt{\Gamma_{1R} \Gamma_{dR}} & \sqrt{\Gamma_{2R} \Gamma_{dR}} & \cdots \\
\sqrt{\Gamma_{dR} \Gamma_{1R}} & \Gamma_{1R} & \sqrt{\Gamma_{2R}\Gamma_{1R}} & \cdots \\
\sqrt{\Gamma_{dR} \Gamma_{2R}} & \sqrt{\Gamma_{1R}\Gamma_{2R}} & \Gamma_{2R} & \cdots \\
\vdots & \vdots & \vdots & \vdots 
\end{array}
\right)\; , \;
\mathbf{\Gamma}^L = \left(  
\begin{array}{cccc}
\Gamma_{dL} & 0  & 0 & \cdots \\
0 & 0 & 0 & \cdots \\
0 & 0 & 0 & \cdots \\
\vdots & \vdots & \vdots & \vdots 
\end{array}
\right)\; , \;
\mathbf{G}^{\alpha} = \left(  
\begin{array}{cccc}
G_{dd}^{\alpha}(\omega)  & G_{d1}^{\alpha}(\omega)  & G_{d2}^{\alpha}(\omega) & \cdots \\[0.3cm]
G_{1d}^{\alpha}(\omega)  & G_{11}^{\alpha}(\omega)  & G_{12}^{\alpha}(\omega) & \cdots \\[0.3cm]
G_{2d}^{\alpha}(\omega)  & G_{21}^{\alpha}(\omega)  & G_{22}^{\alpha}(\omega) & \cdots \\[0.3cm]
\vdots & \vdots & \vdots & \vdots 
\end{array}
\right)\; .
\end{equation}

In this notation, it is clear that the system cannot be proportionally coupled since $\mathbf{\Gamma}^L \neq \lambda \mathbf{\Gamma}^R$ always.

Our task now is to rewrite Eq.\ (\ref{eq:JR_intermediario1}) in terms of dot Green's functions only. Inserting the expressions for the lesser Green's function in Eq.\ (\ref{eq:JR_intermediario1}) we obtain
\begin{eqnarray}
\label{eq:JR_intermediario2}
J_R &=& \frac{i e}{h}\sum_{\sigma}  \int \! d\omega \, \Bigg\{ \Gamma_{dR} 
\Big[ f_R^{}(\omega) [G_{dd}^r(\omega)-G_{dd}^a(\omega)]+  G_{dd}^<(\omega) \Big] 
\nonumber\\
& & +2\sqrt{\Gamma_{dR}} \,\sum_j  \sqrt{\Gamma_{jR}}\Big\{ 
f_R(\omega)\left[G_{jd}^r(\omega)-G_{jd}^a(\omega)\right ]+ \frac{1}{2}\sum_l 
\left[ G^{(0),r}_{jl}(\omega) \widetilde{\Sigma}_{ld}^{R,r}(\omega) + 
G^{(0),a}_{lj}(\omega) \widetilde{\Sigma}_{dl}^{R,a}(\omega)\right]
G_{dd}^{<}(\omega) \nonumber \\ 
&&   + \left[ G^{(0),a}_{lj}(\omega) \widetilde{\Sigma}_{dl}^{R,<}(\omega) 
+ G^{(0),<}_{lj}(\omega) \widetilde{\Sigma}_{dl}^{R,r}(\omega)\right]
G_{dd}^{r}(\omega)+\left[ G^{(0),r}_{jl}(\omega) \widetilde{ 
\Sigma}_{ld}^{R,<}(\omega) + G^{(0),<}_{jl}(\omega) \widetilde{ 
\Sigma}_{ld}^{R,a}(\omega)\right] G_{dd}^{a}(\omega) \Big\} \nonumber \\
&&+ \sum_{ij}  \sqrt{\Gamma_{iR}\Gamma_{jR}}\Big\{ f_R(\omega) 
\left[G_{ij}^r(\omega)-G_{ij}^a(\omega)\right] +{\frac{1}{2} 
\left(G_{ij}^{(0),<}(\omega)+G_{ji}^{(0),<} (\omega)\right)} \nonumber \\
&& 
+\frac{1}{2}\sum_{ll^\prime}\left(G_{il}^{(0),r}(\omega)\tilde\Sigma_{ld}^{R,
r }(\omega)\tilde\Sigma_{dl^\prime}^{R,a}(\omega) G_{l^\prime j}^{(0),a}(\omega) + 
{G_{jl^\prime}^{(0),r}(\omega)\tilde\Sigma_{l^\prime 
d}^{R,r}(\omega)\tilde\Sigma_{dl}^{R,a}(\omega) G_{l i}^{(0),a}(\omega)} 
\right)G_{dd}^{<}(\omega)\nonumber\\
&& + \left[
G_{il}^{(0),r}\widetilde{\Sigma}_{ld}^{R,r}\left(\widetilde{\Sigma}_{
dl^\prime}^{R,r} G_{l^\prime j}^{(0),<} + \widetilde{\Sigma}_{ 
dl^\prime}^{R,<} G_{l^\prime j}^{(0),a} \right) + {G_{jl^\prime}^{(0),r}
\widetilde{\Sigma}_{l^\prime d}^{R,r}(\omega) 
\left(\widetilde{\Sigma}_{dl}^{R,r} G_{li}^{(0),<} + 
\widetilde{\Sigma}_{dl}^{R,<} G_{l i}^{(0),a} \right)}\right]
G_{dd}^{r}(\omega) \nonumber\\
&& + \left[ \left(G_{il}^{(0),r}\tilde\Sigma_{ld}^{R,<} 
+ G_{il}^{(0),<}\widetilde{\Sigma}_{ld}^{R,a} \right)  
\widetilde{\Sigma}_{dl^\prime}^{R,a} G_{l^\prime j}^{(0),a}+ 
{\left(G_{jl^\prime}^{(0),r}\tilde\Sigma_{l^\prime d}^{R,<} 
 + G_{j l^\prime }^{(0),<}\widetilde{\Sigma}_{l^\prime d}^{R,a} \right) 
\widetilde{\Sigma}_{d l}^{R,a} G_{l i}^{(0),a}} \right]G_{dd}^{a}(\omega)
\Big\}\Bigg\}.
\end{eqnarray}
%
%
In the above we need to know that
\begin{align}
G_{ij}^{(0),r}(\omega) = &\, \left[\left[\omega -  H_{\rm cavity} - 
\Sigma^{R,r}(\omega)\right]^{-1}\right]_{ij}, 
\\
G_{ij}^{(0),a}(\omega) = &\, \left[\left[\omega -  H_{\rm cavity} - 
\Sigma^{R,a}(\omega)\right]^{-1}\right]_{ij},
\\
G_{ij}^{(0),<}(\omega) = &\,\sum_{ll'} G_{il}^{(0),r}(\omega)  
\,\Sigma_{ll'}^{R,<}(\omega)\,G_{l'j}^{(0),a}(\omega),
\end{align}
correspond to the cavity Green's functions in the absence of the 
dot.

\subsection{Expressions for the current}
\label{sec:CurrentSimplified}

We now write the results of Eq.\ (\ref{eq:JR_intermediario2}) for the simplified 
model. First note that by assuming that the coupling of all the cavity levels with the right 
reservoir are equal, i.e., $V_{iR}=V_{cR}$ for all $i$ in the cavity. Then $\Sigma_{ij}^R(\omega)=
-i\Gamma_{cR}/2$, $\Sigma_{id}^R(\omega)=-i\sqrt{\Gamma_{dR}\Gamma_{cR}}/2$. 
 %
%
The lesser GF can be written as $\Sigma_{ij}^{R, <}  (\omega)=i f_R(\omega) 
\Gamma_{cR}$ using the fact that $\sum_{{\bf k}}g^<_{{\bf k},R}(\omega)=2 i \pi 
\rho_R f_R(\omega)$.
Assuming also $V_{di}=\Omega$ for all dot-level coupling matrix elements, the self-energies defined 
in Eqs.~ \eqref{eq:TildeSelfEnergies1} to \eqref{eq:TildeSelfEnergies3} can also be simplified,
\begin{align}
\label{eq:TildeSigmaSM}
\widetilde{\Sigma}_{jd}^{R,(a,r)}(\omega) = &\,  \Omega \pm i\sqrt{\Gamma_{dR}\Gamma_{cR}}/2
\\
\label{eq:TildeSigmaLesser}
\widetilde{\Sigma}_{jd}^{R,<}(\omega) = &\, +  i f_R(\omega) \sqrt{\Gamma_{dR}\Gamma_{cR}}.
\end{align}

We can use the method of equations of motion (EOM) to write $J_R$ and $J_R$ in terms of the Green's function for the dot. Using the Eqs.~\eqref{eq:TildeSigmaSM} into Eq.~\eqref{eq:G0ij} 
we write
\be
\label{eq:G0ijSM}
\sum_{ij} G^{(0),r}_{ij}(\omega) = \tilde S(\omega) ,
\ee
with $\tilde S(\omega)=S(\omega)(1 + i S(\omega) \Gamma_{cR}/2)^{-1}$, where 
$S=\sum_{i}(\omega -\ve_i)^{-1}$. The  lesser GF becomes
\be
\label{eq:G0lesserSM}
\sum_{ij} G^{(0),<}_{ij}(\omega) =  i f_R(\omega)  
\Gamma_{cR}\left|\tilde S(\omega)\right|^2 \; .
\ee

From there, we can write the other Green's functions we are going to need in 
terms of dot's GFs:
\begin{equation}
\label{eq:GridSM}
\sum_{i} G^{r}_{id}(\omega) = \tilde S(\omega) \left( \Omega - i \sqrt{\Gamma_{cR} \Gamma_{dR}} /2\right) G^{r}_{dd}(\omega),
\end{equation}
\begin{equation}
\label{eq:GrijSM}
\sum_{ij} G^{r}_{ij}(\omega) = \tilde S(\omega)\Big[1 + \tilde S(\omega) \left(\Omega - i\sqrt{\Gamma_{cR}\Gamma_{dR}}/2\right)^2 G^{r}_{dd}(\omega)  \Big].
\end{equation}

We are now in a position of re-writing Eq.~\eqref{eq:JR_intermediario2}  for this simplified model. 
Let us start with by writing it in terms of the  GFs defined 
above (sums included). Using 
Eqs.~\eqref{eq:ResonanceSelfEnergies-flat},  \eqref{eq:TildeSigmaSM}, and \eqref{eq:TildeSigmaLesser}, 
the Eq.~\eqref{eq:JR_intermediario2} becomes
%
\begin{eqnarray}
\label{eq:JR_simplifiedModel} 
J_R &=& \frac{ i e}{h}\sum_{\sigma}  \int \! d\omega \Bigg\{
\Gamma_{dR} \Big[ f_R^{}(\omega) (G_{dd}^r -G_{dd}^a )+  G_{dd}^<  \Big] 
+2\sqrt{\Gamma_{dR} \Gamma_{cR}} \, \Bigg\{ f_R(\omega) 
\Big[\sum_j G_{jd}^r - \sum_j G_{jd}^a 
\Big] \nonumber \\
&& +  \frac{1}{2}\left[ 
\Big(\sum_{jl} G^{(0),r}_{jl}\Big) \left(\Omega - \frac{i}{2}\sqrt{\Gamma_{dR}\Gamma_{cR}} \right)+
\Big(\sum_{jl} G^{(0),a}_{lj}\Big) \left(\Omega+ \frac{i}{2}\sqrt{\Gamma_{dR}\Gamma_{cR}}\right)
\right] G_{dd}^{<} \nonumber \\ 
&& + \frac{1}{2}\left[
\Big(\sum_{jl} G^{(0),a}_{lj}\Big) \, i f_R^{}(\omega)\sqrt{\Gamma_{dR}\Gamma_{cR}}+
\Big(\sum_{jl} G^{(0),<}_{lj}\Big) \left( \Omega -\frac{i}{2}\sqrt{\Gamma_{dR} \Gamma_{cR}} \right) 
\right] G_{dd}^{r} \nonumber \\
&& + \frac{1}{2}\left[
\Big(\sum_{jl} G^{(0),r}_{jl}\Big)\,  i f_R^{}(\omega)\sqrt{\Gamma_{dR}\Gamma_{cR}}+
\Big(\sum_{jl} G^{(0),<}_{jl}\Big) \left(\Omega+\frac{i}{2}\sqrt{\Gamma_{dR} \Gamma_{cR}}  \right) 
\right] G_{dd}^{a} \Bigg\}\nonumber \\
&&+ \Gamma_{cR} \Bigg\{ f_R(\omega)  \Big(\sum_{ij} G_{ij}^r-\sum_{ij} G_{ij}^a \Big) +
\sum_{ij}G_{ij}^{(0),<}+ \left|\sum_{il} G_{il}^{(0),r}\right|^2 \left( \Omega^2 + \frac{\Gamma_{dR}\Gamma_{cR}}{4} 
\right)G_{dd}^{<} \nonumber \\
&& + \Big(\sum_{il}G_{il}^{(0),r}\Big) \left( \Omega - \frac{i}{2} \sqrt{\Gamma_{dR}\Gamma_{cR}} \right) \left[ \left( \Omega - \frac{i}{2} \sqrt{\Gamma_{dR}\Gamma_{cR}} \right)\sum_{l^\prime j}G_{l^\prime 
j}^{(0),<} + if_R^{}(\omega)\sqrt{\Gamma_{dR} 
\Gamma_{cR}} \sum_{l^\prime j}G_{l^\prime j}^{(0),a}
\right]G_{dd}^{r} \nonumber \\
&& + \Big(\sum_{l^\prime j}G_{l^\prime j}^{(0),a}\Big)
\left(\Omega + \frac{i}{2} \sqrt{\Gamma_{dR}\Gamma_{cR}} \right)
\left[\left( \Omega +\frac{ i}{2} 
\sqrt{\Gamma_{dR}\Gamma_{cR}} \right) \sum_{il}G_{il}^{(0),<}
+ i f_R^{}(\omega)\sqrt{\Gamma_{dR}\Gamma_{cR}} 
\sum_{il}G_{il}^{(0),r} \right]G_{dd}^{a} \Bigg\}\Bigg\}.\nonumber 
\\
\end{eqnarray}

We now substitute  Eqs.\ (\ref{eq:G0ijSM})--(\ref{eq:GrijSM}) into Eq.~\eqref{eq:JR_simplifiedModel} 
and collect the terms in $G^{r}_{dd}$, 
$G^{r}_{dd}$ and $G^{<}_{dd}$. Using the limit $\lim_{\eta \rightarrow 0} 
S(\omega + i\eta)$ (i.e., taking the analytic continuation of $S(\omega)$ to the 
real axis), and after some long but straightforward algebra, we obtain $J_R$ in 
a nice, compact form:
\begin{eqnarray}
\label{eq:JRcompactSM}
J_R =  J^{(0)}_R + \frac{ i e}{h}\sum_{\sigma}  \int \! d\omega \, 
{\tGamma}_R(\omega) \Big\{ f_R^{}(\omega) \left[G_{dd}^r(\omega) 
 - G_{dd}^a(\omega) \right]+ G_{dd}^<(\omega) \Big\}. \quad
\end{eqnarray}

In the equation above,  $J^{(0)}_R$ is a background contribution coming 
from the terms in Eq.\ (\ref{eq:JR_simplifiedModel}) that do not involve dot's 
Green's functions:
\begin{equation}
\label{eq:J0SM}
J^{(0)}_R\!=\!\frac{ i e}{h}\sum_{\sigma}  \int d\omega \Gamma_{cR}  
\sum_{ij}\left\{  f_R^{}(\omega)\left[G_{ij}^{(0),r}(\omega) - G_{ij}^{(0),a}(\omega) 
\right]\right.
\left. +  G_{ij}^{(0),<}(\omega)  \right\}.
\end{equation}
%
In fact, as we will show below, $J^{(0)}_R$ vanish,  explicitly, for 
$\eta\rightarrow 0$. The effective coupling  ${\tGamma}_R(\omega)$ is a 
\textit{real} algebraic function of the parameters, given by:
\begin{equation}
\label{eq:GammatildeR}
{\tGamma}_R(\omega) = \Gamma_{dR} + \Gamma_{cR}  \left|\tilde 
S(\omega)\right|^2 \left( \Omega^2 + \frac{\Gamma_{cR} \Gamma_{dR}}{4}\right)
+\sqrt{\Gamma_{dR}\Gamma_{cR}} \left[\tilde S(\omega) \left( \Omega  
- i \frac{\sqrt{\Gamma_{cR} \Gamma_{dR}}}{2} \right) + \text{H.c.} 
\right] \; .
\end{equation}
From Eq.\ (\ref{eq:JL_intermediario2}) it is straightforward to show that the effective coupling to the left lead is simply ${\tGamma}_L(\omega)\!=\!\Gamma_{dL}$.

These expressions are all exact as long as there are no interactions in the cavity. At this point, a fair question is ``What are the gains by performing such transformations"? The advantage here is that now $J_R$, given by Eq.\ (\ref{eq:JRcompactSM}), is written in terms of dot's Green's functions only, in the same structure as $J_L$ (for which $\Gamma_{iL}=0$) given by Eq.\ \ref{eq:JL_intermediario2}. As shown below, this is a crucial step in the elimination of $G_{dd}^<(\omega)$ in the current expression.


\section{Fluctuation-dissipation theorem}
\label{sec:FluctuationDissipation}

An important consistency check for the expressions given in  the previous section is 
the applicability of the fluctuation-dissipation theorem (FDT). For instance, 
the expression for $J_R$ in Eq.\ (\ref{eq:JRcompactSM}) vanishes in 
equilibrium, when the fluctuation-dissipation  theorem (FDT) applies.

Just a reminder: the FDT states that, for a system  in thermal equilibrium 
with a reservoir described by a Fermi distribution $f_R(\omega)$, the lesser 
Green's function is proportional to the spectral density
\be
G^{<}_{\nu}(\omega) = 2 \pi i f_R(\omega) A_{\nu}(\omega) \; ,
\ee 
where $A_{\nu}(\omega)=(-1/\pi)$Im $G^{r}_{\nu}(\omega)$.

We can put the FDT in terms of retarded and advanced  Green's functions. 
Using  
$G^{a}_{\nu}(\omega)=(G^{r}_{\nu}(\omega))^*$, the FDT implies
\be
\label{eq:FDtheorem}
G^{<}_{\nu}(\omega) = -f_R(\omega) [G^{r}_{\nu}(\omega) - 
G^{a}_{\nu}(\omega)] \; .
\ee 

This is important as a consistency check for the current  calculations. 
Applying Eq.\ (\ref{eq:FDtheorem}), the current to/from a single lead should 
vanish (which is the correct result in equilibrium). This can be readily 
verified, for instance, for $J^{(0)}_R$ defined in Eq.~\eqref{eq:J0SM} and for 
$J_L$ [Eq.~\eqref{eq:JL_intermediario2}].

In fact, this consistency check can be applied to each of  the three terms 
in Eq.\ (\ref{eq:JR_intermediario1}) by verifying that the  FDT is satisfied 
for each of the Green's functions involved. 
Note that the first term in in Eq.\ (\ref{eq:JR_intermediario1})  involves 
diagonal (dot) GFs and is clearly consistent with the FDT: it vanishes if 
$G^{<}_{dd}(\omega) = -f_R(\omega) [G^{r}_{dd}(\omega) - 
G^{a}_{dd}(\omega)]$.

The second term involves non-diagonal Green´s functions.  We can then 
explicitly show that
\begin{eqnarray}
\frac{1}{2}\sum_j G^{<}_{jd} + G^{<}_{dj} = -f_R(\omega)  \left[ \sum_j 
G^{r}_{jd} - G^{a}_{jd} \right] \; .
\end{eqnarray}

The right-hand side of the above expression can be easily  calculated using Eq.~\eqref{eq:GridSM}. 
Using the short-hand notations
\begin{eqnarray}
\tilde{S} & \equiv & \frac{S(\omega)}{1 + i S(\omega) \Gamma_{cR}/2 } \\[0.3cm]
\tilde{\Omega} & \equiv & \Omega + i \sqrt{ \Gamma_{cR} \Gamma_{dR}}/2 \; ,
\end{eqnarray}
we have
\begin{equation}
\label{eq:rhsFDT}
 -f_R(\omega)  \sum_j \left( G^{r}_{jd} - G^{a}_{jd} \right)  =
 -f_R(\omega) \left[ \tilde{S} \; (\tilde{\Omega})^* \; G^{r}_{dd}-(\tilde{S})^* \tilde{\Omega} \; G^{a}_{dd} \right].
\end{equation}

Thus, the left-hand side can be calculated with the help of  Eqs.~(\ref{eq:Gidra}) and (\ref{eq:Gdi_lesser}) 
and from Eqs.~(\ref{eq:TildeSigmaSM}) to (\ref{eq:G0ijSM}), giving
\begin{equation}
 \label{eq:lhsFDT}
\sum_j G^{<}_{jd} + G^{<}_{dj} = \left( \tilde{S}\tilde{\Omega}^* + \tilde{S}^* \; \tilde{\Omega}\right) G^{<}_{dd} \\
+ i f_R(\omega) \left[ \left( \Gamma_{cR} |\tilde{S}|^2 \tilde{\Omega}^* + \sqrt{ \Gamma_{cR} \Gamma_{dR} }\; \tilde{S}^* \right) G^{r}_{dd} 
+\left( \Gamma_{cR} |\tilde{S}|^2 \tilde{\Omega}  + \sqrt{\Gamma_{cR} \Gamma_{dR} }\; \tilde{S} \right) G^{a}_{dd} \right].
\end{equation}


Now, using $G^{<}_{dd}(\omega) = -f_R(\omega) [G^{r}_{dd}(\omega) - 
G^{a}_{dd}(\omega)]$, Eq.\ (\ref{eq:lhsFDT}) reduces to Eq.\ 
(\ref{eq:rhsFDT}). 
In order to show that, we take the limit $S(\omega)=\lim_{\eta 
\rightarrow 
0}  S(\omega + i\eta)$ and then use the following properties:
\begin{eqnarray}
(\tilde{S})^*  & = & \tilde{S} +  i \Gamma_{cR} |\tilde{S}|^2 \\[0.3cm]
\tilde{\Omega} & = & (\tilde{\Omega})^*  +  i \sqrt{ \Gamma_{cR} 
\Gamma_{dR}} \; .
\end{eqnarray}
A similar calculation can be done to show that the third term in  Eq.\ 
(\ref{eq:JR_intermediario1}) also satisfies the FDT.

\section{\texorpdfstring{Meir-Wingreen-like elimination of $G^<$}{Meir-Wingreen-like elimination of G-lesser}}
\label{app:MW-like}

Let us consider the current formula for a single-resonance QD \cite{Supp::MeirWingreen92}
\begin{equation}
\label{eq:MeirWingreen}
J_{R(L)} \equiv \int d\omega \, I_{R(L)} (\omega),
\end{equation}
\begin{equation}
 I_{R(L)} (\omega) = \frac{ie}{h} \tGamma_{L(R)}(\omega) \Big\{ G^<(\omega) + f_{L(R)}(\omega) \left[ G^r(\omega) - G^a(\omega) \right]\Big\}. 
\end{equation}

In the steady state, charge conservation implies that $J_L=-J_R$, hence
\be 
J_L = \frac{J_L - J_R}{2}
\ee
or, in general $J_L = xJ_L - (1-x) J_R$, where $x$ is arbitrary.

We stress that $J_L=-J_R$ is the same as
\be
 \int d\omega \, I_{L} (\omega) = -\int d\omega \, I_{R} (\omega),
\ee
which {\it does not} mean that  $I_{L} (\omega)=-I_{R} (\omega)$ for a given energy $\omega$.

Let us restrict ourselves to the linear response regime and write
\begin{align}
\label{eq:linear_response}
G^<(\omega) = &\, G^<_{\rm eq}(\omega) + \frac{\partial G^<}{\partial  \mu} \Delta \mu 
+ O(\Delta \mu^2) 
\\
f_{L(R)}(\omega) = &\, f_0(\omega) \pm \frac{1}{2}\frac{\partial  f_0}{\partial \mu} 
\Delta \mu  + O(\Delta \mu^2) .
\end{align}
We recall that the fluctuation-dissipation theorem gives
\[
G^<_{\rm eq}(\omega) = -f_0(\omega) \left[G^r(\omega) - G^a(\omega) \right],
\]
allowing us to write the current $J_{L(R)}$, Eq.~\eqref{eq:MeirWingreen}, as
\begin{equation}
J_{L(R)} = \frac{ie}{h} \Delta \mu \int d\omega \, \tGamma_{L(R)}(\omega) \Bigg\{ 
\frac{\partial G^<}{\partial \mu}  \mp \frac{1}{2}  \frac{\partial f_0}{\partial \omega}
\Big[ G^r(\omega) - G^a(\omega) \Big]\Bigg\},
\end{equation}
where $\mp$ refer to the sign of chemical potential offset of $L$ and $R$ terminals with
respect to the Fermi energy.
Affleck and collaborators \cite{Supp::Affleck2013} claim that   $\partial G^</\partial 
\mu$ is expected to have the form $(-\partial f_0/\partial \omega) \Pi(\omega)$, where 
(in general) $\Pi(\omega)$ has a smooth energy dependence on the scale of $kT$. For 
now, we assume this is true.

Let us assume that $\tGamma_{L(R)}(\omega)$ varies slowly with $\ve$ over energies 
scales of the order of $kT$, which is a condition met  in almost all situations 
of interest. In this scenario it is safe to approximate
\begin{equation}
J_{L(R)} \approx \frac{ie}{h} \Delta \mu \,\tGamma_{L(R)}(\ve_F)  \int d\omega \,  
\Bigg\{ \frac{\partial G^<}{\partial \mu} \pm \frac{1}{2} \left(- 
\frac{\partial f_0}{\partial \omega}\right)  \Big[ G^r(\omega) - G^a(\omega) 
\Big]\Bigg\}.
\end{equation}
We now use the general relation $J_L = xJ_L - (1-x) J_R$ to write
%
\begin{align}
J_{L} \approx \frac{ie}{h} \Delta \mu 
\Bigg[ x \,\tGamma_{L}(\ve_F) &\! \int\! d\omega \,  \left\{ 
\frac{\partial G^<}{\partial \mu} + \frac{1}{2} \left(-  \frac{\partial 
f_0}{\partial \omega}\right)  \Big[ G^r(\omega) - G^a(\omega) \Big]\right\}
\nonumber\\
- (1-x)\tGamma_{R}(\ve_F) &\int\! d\omega \,  \left\{ 
\frac{\partial G^<}{\partial \mu} - \frac{1}{2} \left(-  \frac{\partial 
f_0}{\partial \omega}\right)  \Big[ G^r(\omega) - G^a(\omega) \Big]\right\}\Bigg].
\end{align}
%
To eliminate the $G^<$ term one needs $x\tGamma_L - (1-x) \tGamma_R=0$, yielding 
$x=\tGamma_R/(\tGamma_L+ \tGamma_L)$. Hence
\begin{equation}
\label{Wingreen_like}
J_{L} \approx \frac{ie}{h} \Delta \mu \frac{\tGamma_{L}(\ve_F) \tGamma_{R}(\ve_F)}
{\tGamma_{L}(\ve_F) + \tGamma_{R}(\ve_F)} 
  \times \int\! d\omega \,  \left(- \frac{\partial f_0}{\partial \omega}\right)  
\Big[ G^r(\omega) - G^a(\omega) \Big].
\end{equation}
This expression is the same as the one obtained by Meir  and Wingreen 
\cite{Supp::MeirWingreen92} using the proportional coupling trick, namely, by assuming 
that $\tGamma_L(\omega) = \lambda \tGamma_R(\omega)$, where 
$\lambda$ does not depend on energy.

From the expression for the current [Eq.~\eqref{Wingreen_like}]  we can readily 
derive the corresponding expression for the conductance through the system:
\begin{equation}
\label{eq:Conductance}
G = \frac{{2}e^2}{\hbar} \frac{\tGamma_{L}(\ve_F) {\tGamma}_R(\ve_F)}
{\tGamma_{L}(\ve_F) + {\tGamma}_R(\ve_F)} 
 \int\! d\omega \,  \left(- \frac{\partial f_0}{\partial \omega}\right)  
A_{d} (\omega),
\end{equation}
where $f_0$
written in terms of the dot spectral density $A_{d} (\ve,T)=(-1/\pi)  \mbox{Im 
}G_{dd}^r (\omega)$ at temperature T that can be calculated with NRG.

For the cavity system studied in this work, we have ${\tGamma}_L(\omega)\!=\!\Gamma_{dL}$ and ${\tGamma}_R(\omega)$ is given by Eq.\ (\ref{eq:GammatildeR}). We note, however, that this approach is very generic and can be applied to a large class of systems with arbitrarily complex geometries and for which interactions are restricted to a single level.

%


\begin{thebibliography}{14}
\bibitem [{Lan()}]{Supp::Langreth}%
  \BibitemOpen
  \href@noop {} {}\bibinfo {note} {The Langreth rules
  \cite{Langreth:Book::1976} we have employed here can be written as
  \begin{eqnarray*} C^r &=& A^r B^r \Rightarrow C^< = A^r B^< + A^< B^a \\ D^r
  &=& A^r B^r C^r \Rightarrow D^< = A^r B^r C^< + A^r B^< C^a + A^<B^a C^a,
  \end{eqnarray*} where $A$, $B$ and $C$ are matrices.}\BibitemShut {Stop}%
\end{thebibliography}
\end{document}